\documentclass{emulateapj}

\usepackage{apjfonts}

\begin{document}

\shorttitle{MODELING THE MAGNETIC FIELD IN THE GALACTIC DISK}
\shortauthors{VAN ECK ET AL.}

\slugcomment{Accepted for publication in ApJ; December 13, 2010}

\title{Modeling the Magnetic Field in the Galactic Disk using New Rotation Measure Observations from the Very Large Array}
\author{C.L. Van Eck$^1$, J.C. Brown$^1$, J.M. Stil$^1$, K. Rae$^1$, S.A. Mao$^{2,3}$, B.M. Gaensler$^4$, \\
A. Shukurov$^5$, A.R. Taylor$^1$, M. Haverkorn$^{6,7}$, P.P. Kronberg$^{8,9}$, N.M. McClure-Griffiths$^3$}

\vspace{5mm}

\affil{\footnotesize{1. Dept. Physics \& Astronomy, University of Calgary, T2N 1N4, Canada}}
\affil{\footnotesize{2.Harvard-Smithsonian Center for Astrophysics, Cambridge, MA 02138, USA}}
\affil{\footnotesize{3. Australia Telescope National Facility, CSIRO Astronomy and Space Science, PO Box 76, Epping, NSW 1710, Australia}}
\affil{\footnotesize{4. Sydney Institute for Astronomy, School of Physics, The University of Sydney, NSW 2006, Australia}}
\affil{\footnotesize{5. School of Mathematics and Statistics, University of Newcastle, Newcastle upon Tyne, NE1 7RU, UK}}
\affil{\footnotesize{6. ASTRON, Oude Hoogeveensedijk 4, 7991 PD Dwingeloo, The Netherlands}}
\affil{\footnotesize{7. Leiden Observatory, Leiden University, P.O. Box 9513, 2300 RA Leiden, The Netherlands}}
\affil{\footnotesize{8. Department of Physics, University of Toronto, 60 St. George Street, Toronto M5S 1A7, Canada}}
\affil{\footnotesize{9. Los Alamos National laboratory, M.S. T006, Los Alamos NM 87545 USA  }}

\email{clvaneck@ucalgary.ca; jocat@ras.ucalgary.ca}


\begin{abstract}
We have determined 194 Faraday rotation measures (RMs)  of polarized extragalactic radio sources using new, multi-channel polarization observations at frequencies around 1.4 GHz from the  Very Large Array (VLA)  in the Galactic plane at $17^\circ \leq l \leq 63^\circ$ and $205^\circ \leq l \leq 253^\circ$.  This catalog fills in gaps in the RM coverage of the Galactic plane between the  Canadian Galactic Plane Survey and Southern Galactic Plane Survey.  Using this catalog we have tested the validity of recently-proposed axisymmetric and bisymmetric models of the large-scale (or regular) Galactic magnetic field, and found that of the existing models we tested, an axisymmetric spiral model with reversals occurring in rings (as opposed to  along spiral arms) best matched our observations.  Building on this, we have performed our own modeling, using RMs from both extragalactic sources and pulsars. By developing independent models for the magnetic field in the outer and inner Galaxy, we conclude that in the inner Galaxy, the magnetic field closely follows the spiral arms, while in the outer Galaxy, the field is consistent with being purely azimuthal.
 Furthermore, the models contain no reversals in the outer Galaxy, and together seem to suggest the existence of a single reversed region that spirals out from the Galactic center.
\end{abstract}

\keywords{Galaxy: structure --- ISM: magnetic fields --- polarization}

\section{Introduction}

While the importance of the Galactic magnetic field is undisputed -- ranging from star formation
to large-scale galactic dynamics --  much remains unknown
about how the field is generated or how it is evolving.  The only way to address  these questions is to
fully understand the present overall structure of the field, which is an essential constraint to proposed evolutionary models of the field.

One observation central to the study of the Galactic magnetism is that of the Faraday rotation measure (RM).
As a linearly polarized electromagnetic wave propagates through a region containing free thermal electrons and
a magnetic field, such as the interstellar medium,  the plane of polarization will rotate through the
process known as Faraday rotation.  The amount of rotation, $\Delta \phi$ [rad],  experienced by a wave of a given
wavelength, $\lambda$ [m], is given by
\begin{equation}
\Delta \phi  = 0.812 \lambda^2 \int_{\rm source}^{\rm receiver} n_e \mathbf{B \cdot} {\rm{d}}\mathbf{l}, \label{eqn:RM}
\end{equation}
where  $n_e$ [cm$^{-3}$] is the electron density, $\bf{B} [\mu$G] is the magnetic field, d$\bf{l}$ [pc] is the path length element.  
Assuming that the polarization angle at the source, $\phi_\circ$, is the same 
for all emitted wavelengths, that Faraday rotation is the only mechanism acting on the polarization angle, and that the source of interest is the only source contributing polarized emission along the line of sight, then the detected polarization angle, $\phi$, is  a linear function of the square of 
the wavelength through the relationship
\begin{equation}
\phi = \phi_0 +  \lambda^2 {\rm RM},
\end{equation}
where RM is the Rotation Measure, defined from equation \ref{eqn:RM}. 
As a result, measuring the polarization angle at several wavelengths for a given source can provide a simple 
determination of the rotation measure for that line of sight,  provided that there is no internal Faraday dispersion by
turbulent magnetic fields \citep[see e.g.][]{Sokoloff98}. 

If there exist multiple polarized emitting regions along a line of sight, for example, from different regions within the
primary source, or from different regions within the interstellar medium of the Galaxy itself,  or turbulent cells with random field directions,  each emitting region will
 have a different RM contributing to the final detected RM, otherwise known as  {\it{RM components}}.  A technique
developed by \citet{RMSynth}, known as RM Synthesis, is able to extract the RM components using Fourier transforms.
 However, there remains the difficulty of determining where along the line of sight the components originate, and how
  the components `add' is highly dependent on the instrumentation used to detect the signals.  Therefore, it is more
   straightforward to use single component RM sources to probe the magnetic field within the Galaxy.  The two sources
    most often used are pulsars (within the Galaxy) 
and compact extragalactic sources (EGS).  We note that the higher the angular density of the observed probes, the greater the
capacity to identify and separate the ordered and random field components.

For the purposes of modeling, the Galactic magnetic field is often, and somewhat arbitrarily, divided into
two components: a large-scale or regular component, {\bf B}$_u$, with spatial scales on the order 
of a few kpc,  and a turbulent or random component 
{\bf B}$_r$, with spatial scales on the order of tens of pc  \citep{Ruzmaikin88},  with significant different 
spatial scales observed between and within the spiral arms (Haverkorn et al. 2006, 2008). \nocite{mh06, HB08}
 Furthermore, the regular component is 
observed to be concentrated in the disk \citep{sk80}, with a dominant azimuthal component,
some radial component (thus indicating a spiral field), and a weak
vertical or $z$ component \citep{Ann10}.

Additionally, recent work has focused on developing and testing competing models, and determining the existence of large scale reversals in the magnetic field.  Magnetic field reversals occur where the magnetic field direction completely reverses over a short change in radius and/or azimuth  within the disk of the Galaxy.  The number of reversals depends on the interpretation of the existing RM data and is presently a very controversial subject \citep[e.g.][]{bt01, Weisberg04,Vallee05, Han06,Brown07, Sun08, Vallee08, Men08, Ronnie, Kronberg, Katgert}.  Models of the Galactic magnetic field are generally classified as axisymmetric spiral (ASS), which requires symmetry under the same rotation of the disk by $\pi$ around the Galactic center, bisymmetric spiral (BSS), which requires anti-symmetry under rotation by $\pi$, or mixed spiral structure (MSS), which contain both ASS and BSS components \citep{Beck96}.  Most models  are made to follow the spiral arm structure of the Galaxy since an approximate alignment of the regular magnetic fields and spiral arms is commonly observed
in external galaxies \citep[e.g.][]{Beck07}. 

For all of these models, sufficient numbers of low-latitude, high quality RM data in key regions have been lacking. While the recent catalog of \citet{Caleb} significantly increases the number of published EGS RM sources across the entire sky, those in the plane lack adequate reliability for modeling the magnetic field in the disk (see section \ref{sec:observations}).

In this paper, we present our new low-latitude EGS RM catalog derived from new observations from the Very Large Array (VLA), filling in gaps in Galactic plane coverage in quadrant 1 (Q1; $0^\circ < l < 90^\circ$) and quadrant 3 (Q3; $180^\circ < l < 270^\circ$).  We use these data to further discriminate between three popular models investigated by \citet{Sun08}.
We then combine these data with previous observations  to develop magnetic field models for three separate sectors of the
Galactic disk and find some remarkable consistencies between the sectors, suggesting global features of the field.

\section{Observations}\label{sec:observations}

\begin{figure}[htbp]
\plotone{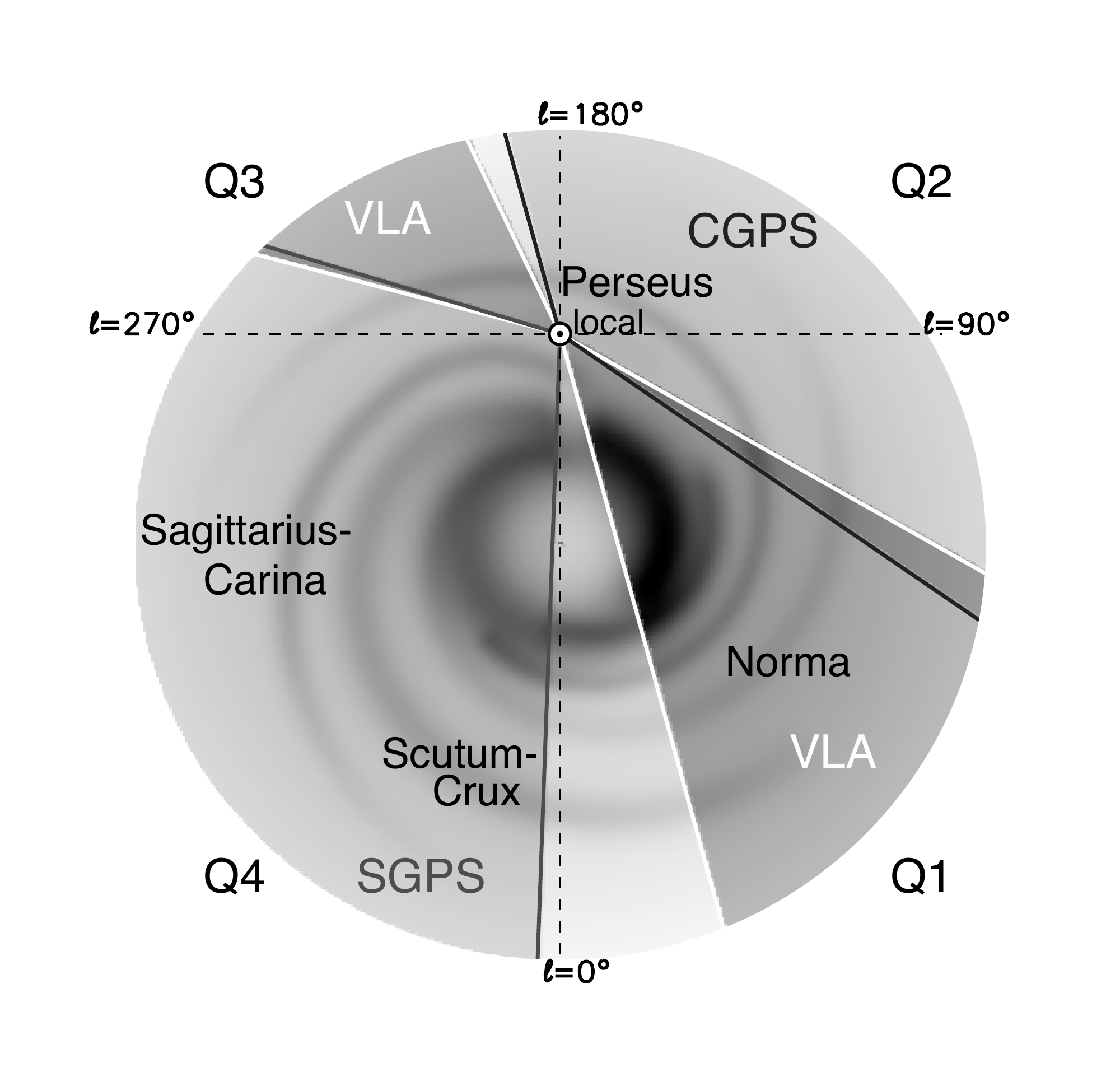}
\caption{View of the Milky Way from above the north Galactic pole illustrating the main survey regions of extragalactic rotation measures used in this paper.  The grey scale background is the electron density distribution model of Cordes and Lazio (2002).  The dark lines are the boundaries of the regions observed by the CGPS and SGPS, while the white lines denote the 2 areas targeted by the VLA data used for this project.  The dashed lines show the delineations between
the Galactic quadrants (Q1 - Q4). }
\label{fig:coverage}
\end{figure}

Two recent projects have produced catalogues for several hundred EGS RMs in the plane of the Galaxy: the Canadian Galactic Plane Survey \citep[CGPS;][]{CGPS, CGPSRMs} and the Southern Galactic Plane Survey \citep[SGPS;][]{Haverkorn06, Brown07}.  While covering a significant fraction of the Galactic disk, these surveys left two gaps in the EGS RM coverage of the Galactic plane as shown in Figure \ref{fig:coverage}.

\nocite{NE2001}

In June and July of 2008, we carried out  48 hours of observing with the VLA under the program AM959 to fill in the gaps in EGS RMs between the CGPS and SGPS. Observations done on June 8 and June 20 were carried out in DnC configuration where we gave priority to low-declination sources, while the remaining observations  (carried out on June 29, July 4, July 5, July 10, and July 12) were in D configuration.   Our observations  in Q1 were confined to $17^\circ \leq l \leq 43^\circ$, $-3^\circ \leq b \leq 3^\circ$ and  $43^\circ < l \leq 63^\circ$, $-4^\circ \leq b \leq 4^\circ$ and in Q3 with $205^\circ \leq l \leq 253^\circ$, $-5^\circ \leq b \leq 5^\circ$.  We increased the latitude coverage for higher longitudes to maximize the number of sources we had, while still maintaining lines-of-sight  largely confined within the Galactic disk under the assumptions of a Galactic radius of 20 kpc and a warm ionized medium scale height of 1.8 kpc \citep{Gaensler}.

The sources observed were selected from the NRAO-VLA Sky Survey \citep[NVSS;][]{NVSS} with the following criteria: 1) they were unresolved, having an NVSS fitted major axis of less than 60 arcseconds; 2) their linearly polarized flux as given in the 
published NVSS catalogue was greater than 2 mJy (bias corrected); 3) they had a minimum fractional polarization of 2\% in Q3 or 1\% in Q1.  We used a different minimum fractional polarization in Q1 than in Q3  to minimize selection against high RMs as a result of bandwidth depolarization in the NVSS.  Using these criteria, we observed a total of 486 sources.

We observed each source in spectral-line mode, using an integration time of at least 2 minutes
in two separate 25 MHz bands, each with 7 channels, giving 14 wavelengths and 14 corresponding polarization angles.   For our first observations of  76 sources (June 8),  we centered the two bands at 1365 MHz and 1515 MHz.  The 1515 MHz band was unusable due to radio frequency interference (RFI), so the sources from these observations were discarded. For all subsequent sources, we used bands centered at 1365 MHz and 1485 MHz.

When possible, we observed a primary flux calibrator (3C286 or 3C138) in both observing bands at the beginning and the end of
an observation run.  For the duration of a given run, we observed sources in groups of 15-18  sources in one band within a 1 hour window.  Every hour, we visited a phase calibrator that was usually within 10 degrees of the target group.  We then repeated the
observations of the target group in the second band (i.e. band A: phase calibrator $\rightarrow$ band A: target group:  $\rightarrow$  band A: phase calibrator $\rightarrow$ band B: phase calibrator  $\rightarrow$
band B: target group  $\rightarrow$ band B: phase calibrator $\rightarrow$ {\it repeat}).

Data reduction was carried out using AIPS software.  The AIPS task IMAGR was used to do imaging and cleaning, which was
performed down to the noise level for each source (between 0.6 mJy and 1.1 mJy).  CLEAN boxes were defined tightly around the target source to prevent cleaning of sidelobes of the dirty beam. Baselines shorter than 1 k$\lambda$ were excluded to reduce extended emission (equalling about 30\% of all baselines).  No additional weighting was used in the UV plane.
Polarization calibration was done using the standard calibration method as described
in the AIPS manual, using 3C286 as a polarization calibrator.  We note that polarization calibration in AIPS can only be done on
one channel at a time.  Thus, each of the 14 channels was calibrated and imaged separately, using 10\arcsec \ pixels and
a 50\arcsec \ clean beam.  The polarization angle and polarized intensity for each channel (i.e. at each wavelength)  were then extracted from the source pixel with the greatest channel-averaged polarized intensity  (determined by averaging the polarized intensity from the individual channels). 

\begin{figure}[htbp]
\plotone{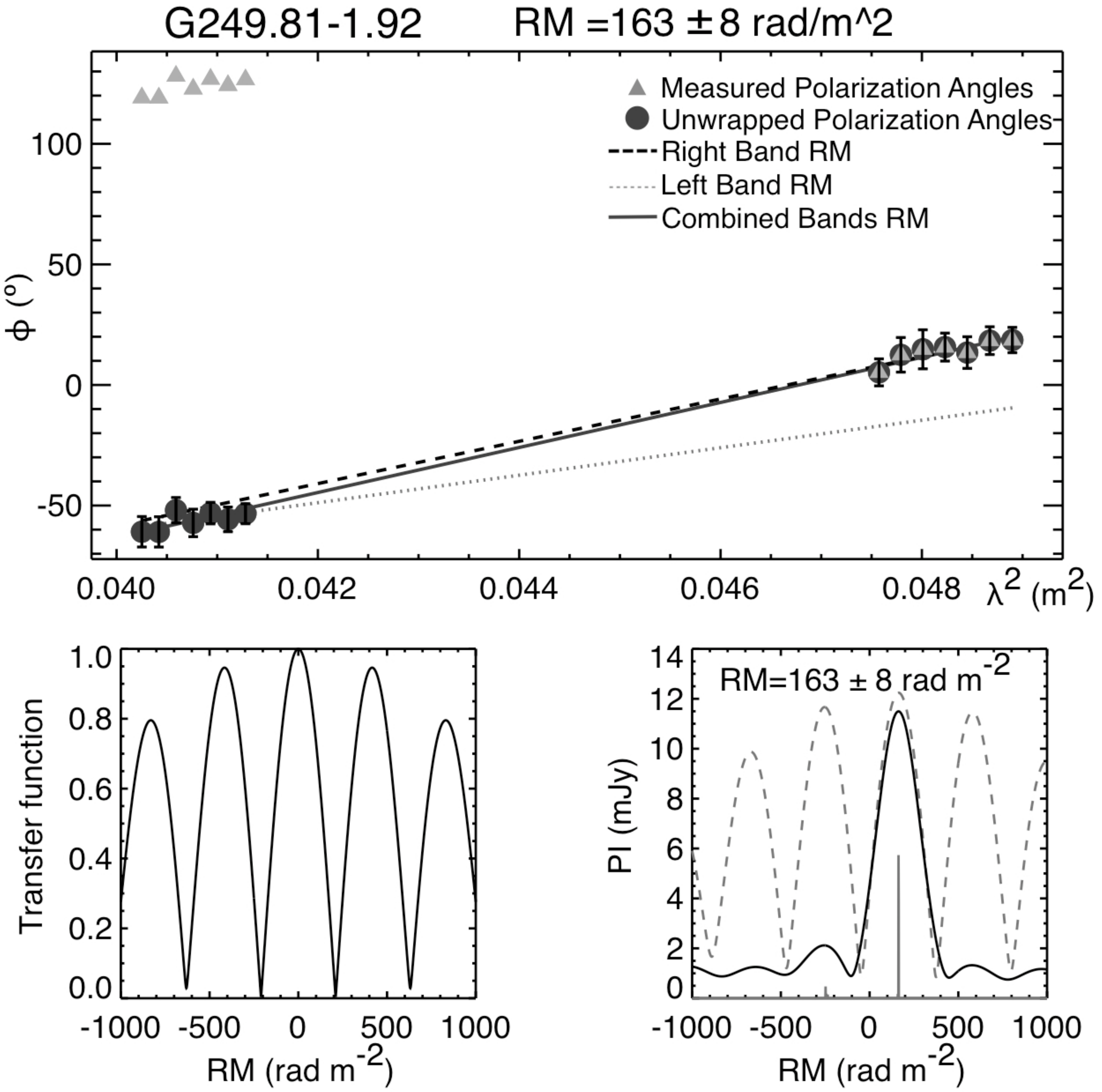}
\caption{Determining rotation measures for the VLA observations. Top: A demonstration of the polarisation angle $n\pi$ determination method for rotation measure calculations as described in the text.  The triangles are the observed angles, circles are the adjusted angles. The dashed (dotted) lines are the extrapolations of the right (left) band, and are used to determine the inter-band $n\pi$ adjustment needed to produce the best fit.  The solid line is the rotation measure determined using both bands after applying the algorithm.  Bottom left: The rotation measure transfer function for RM synthesis resulting from the 14-channel sampling.  Bottom right: The RM Synthesis result for the same source.  The light dotted line is the raw spectrum, the grey solid line is the RM CLEAN component spectrum, and the solid black line is the final CLEANed spectrum.}
\label{fig:unwrap}
\end{figure}

 From these polarization angles, the RM is calculated using a least-squares fit to the polarization angle versus wavelength-squared plot, as reflected by equation (2). Since polarization angles are only defined between    0 and $\pi$,  
 all polarization angle measurements are subject to an issue of `$n\pi$ ambiguity'.  To address this issue within our data, we have used the algorithm used by \citet{CGPSRMs} to determine the relative angles between channels within each band separately.  To determine the  $n\pi$ between the two 25 MHz bands observed,  best fit lines were determined separately for each band and extrapolated out to the other band.  If the determined value for $n$ as determined by the separate bands
 was different, the source was discarded.  
 The bands were then adjusted by the appropriate $n\pi$, and the RM was determined from the slope of the best fit line using both bands together (see Figure
\ref{fig:unwrap}). The error in RM was determined as the error in slope of the 2-band best fit line.

In order to be considered a reliable RM value, the source 
had to have a `probability-of-fit' for the least-squares fit  of polarization angle versus $\lambda^2$ be greater than
the standard value of 10\%.  In addition, the source also had to pass two more tests that were designed 
to confirm that the $n\pi$ algorithm worked consistently and that the source has a {\it single} RM value (or, alternatively, a consistent, dominant RM component). First, we confirmed that the 4 pixels adjacent to the central pixel (above, below, and to the sides) had similar RMs.  Sources with the mean RM of the 4 adjacent pixels greater  than 2$\sigma$ different than the central pixel RM value were rejected.  Second, we confirmed that the fractional polarization (linearly polarized intensity divided by Stokes I) was
constant with $\lambda^2$.   While the presence of multiple RM components may still combine to produce 
linear behavior in polarization angle versus $\lambda^2$ over small wavelength ranges (thereby leading to incorrect
calculated RMs), multiple components are also expected to produce nonlinear behavior in polarized {\it{intensity}} versus $\lambda^2$ \citep{GR84, Farnsworth10}.  By contrast, a single (or at least, dominant) RM component is
expected to have polarized intensity versus $\lambda^2$ be roughly constant.
 Therefore, we used a second `probability-of-fit' test (also at the 10\% level) of the fractional polarization values versus $\lambda^2$ against a line with zero slope and offset equal to the mean fractional polarization.  Sources failing this test were also discarded.

\begin{figure}[ht]
\begin{minipage}{\textwidth}
\plotone{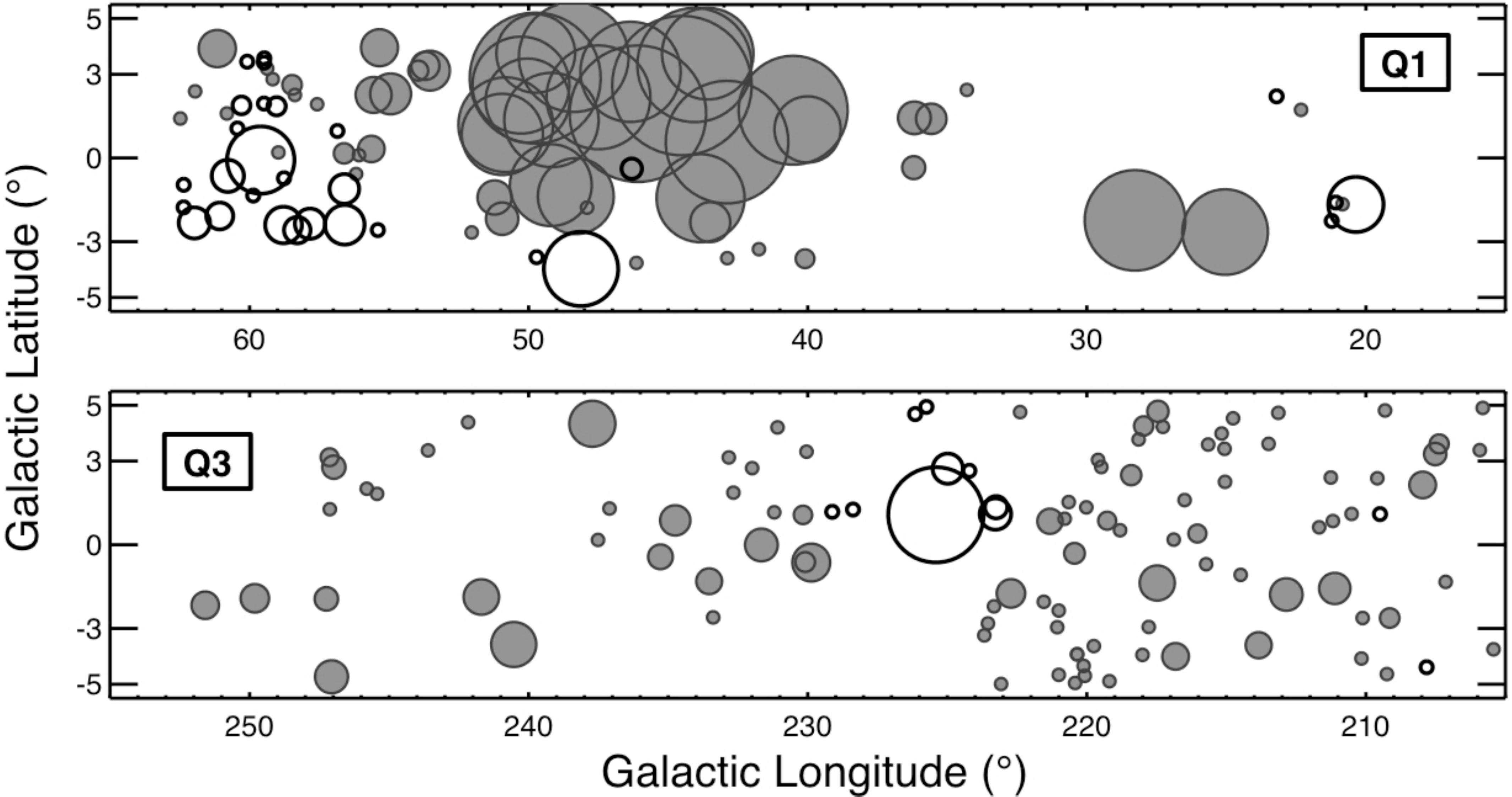}
\caption{Rotation measure sources from Table \ref{table:1}.  Grey filled sources indicate positive rotation measure and black open
symbols indicate negative rotation measures; diameters of symbols are linearly proportional to the magnitude of  RM truncated between
 100 and 600 rad m$^{-2}$ so that sources with $|$RM$| < $ 100 rad m$^{-2}$ are set to 100 rad m$^{-2}$, and those with
 $|$RM$| > $ 600 rad m$^{-2}$ are set to 600 rad m$^{-2}$. The top panel represents sources from Galactic quadrant 1 (Q1), while the bottom panel
 represents sources from Galactic quadrant 3 (Q3).  }
\label{fig:circles}
\end{minipage}
\end{figure}

Using this method, we determined reliable RMs for 194 sources as documented in Table \ref{table:1}, and illustrated in Figure \ref{fig:circles}.
As a consistency check of this catalog, we have compared our RMs to the following two determinations of RMs for the same sources.    In addition to the linear fitting method described above, we also calculated RMs from the same data using RM synthesis, as outlined by \citet{RMSynth}, which employs Fourier transforms to determine the RM.  The resulting RM spectra were processed with the RM CLEAN algorithm, as outlined in \citet{Heald09}.  The values found with the synthesis and linear fitting methods are in complete agreement, with the variation between the two methods much less than the stated error in RM.  The RM synthesis results are included in Table 
\ref{table:1}.  

\begin{figure}[hb]
\plotone{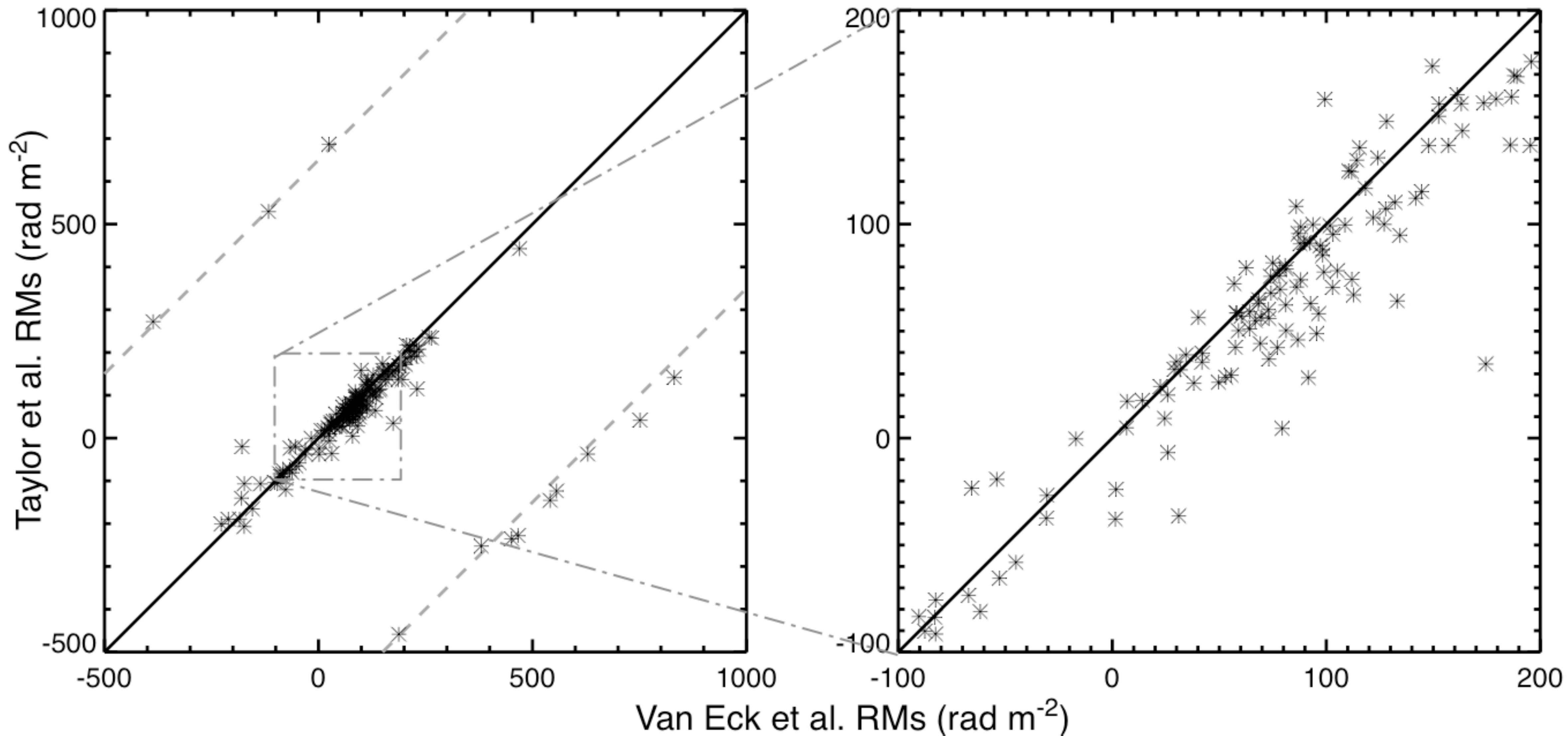}
\caption{Comparison of NVSS RMs calculated by \citet{Caleb}  to our new VLA RMs presented here.  The solid line is the 1:1 line corresponding to complete agreement.  In the left panel, the dashed lines are parallel to the 1:1 line with the $\pm$650 rad m$^{-2}$ offset that a $\pm\pi$ RM ambiguity error would add to an RM determined by the \citet{Caleb} algorithm (as they discuss, the value of the shift from a 1$\pi$ ambiguity error is determined by the two wavelengths for which polarization information is available for the NVSS catalog). Roughly 5\% of the corresponding sources have a $\pm$650 rad m$^{-2}$ offset. Given that a 1$\pi$ RM ambiguity error in a source determined from our algorithm would give a significantly larger offset, we attribute the offsets of this limited number of sources to unresolved ambiguities in those specific \citet{Caleb} sources. The right panel is an enlargement of  part of the left panel,  showing the scatter of the points about the 1:1 curve. There is a slight systematic shift of the \citet{Caleb} RMs towards more negative values, with a 23 rad m$^{-2}$ standard deviation of the difference between the two.}
\label{fig:NVSS}
\end{figure}

The second comparison was with the RM catalog produced directly from the original NVSS observations by \citet{Caleb}, where RMs are calculated using observations at 1364.9 and 1435.1 MHz.  Their method used two polarization angle measurements, combined with depolarization information to solve the half-rotation ambiguity.  
Figure \ref{fig:NVSS} is a comparison of RMs of sources for which there are values from both surveys. It is clear that some of the sources have RMs determined by the two techniques that differ by $\sim$650 rad m$^{-2}$.  A 1$\pi$ ambiguity introduces a $\sim$650 rad m$^{-2}$ offset in a RM determined by \citet{Caleb}.  For our data,  
however,  a 1$\pi$ ambiguity would require an actual RM of greater than 10,000 rad m$^{-2}$.  
Consequently, it is certain that the offset sources in Figure \ref{fig:NVSS} reflect the 1$\pi$ ambiguity of the \citet{Caleb} algorithm.   
The linear correlation coefficient for all 146 matched sources is 0.20, but this rises to 0.96 when the 13 offset sources are removed from the calculation.  This correlation is much higher than that found by Mao et al. (2010; 0.39 in the North, and 0.36 in the South), \nocite{Ann10} likely because our data includes a much larger range of RMs which reduces the effect of random errors.
As well as the $\sim$650 rad m$^{-2}$ offset for a few sources, there are additional differences between our RM values and those of \citet{Caleb}.
As demonstrated in Figure \ref{fig:NVSS},  their values are systematically lower than ours, with a mean shift of 10 rad m$^{-2}$. Further,  the standard deviation of the differences between their values and ours (neglecting the 1$\pi$ shifts and differences between sources with large RMs) is 23 rad m$^{-2}$. Given that we are working in the part of the sky, namely the Galactic disk, where RMs vary significantly over relatively short angular distances, and the fact that we systematically resolve n$\pi$ ambiguities more reliably than their technique does in the same part of the sky, there is a significant advantage to using our RM values in our effort to infer magnetic field structure in the plane of the Galaxy. The systematic shift of the RMs inferred from the two techniques, and the variance between the two, undoubtedly indicate errors in one or both of the methods. These errors are worth investigating further, but because they are almost certainly small relative to the real large scale trends in RM in the Galactic plane (the RMs vary systematically by hundreds of rad m$^{-2}$ but the shift and variance are both on the order of 10 rad 
m$^{-2}$) they are not important for this study.

\section{Observational Discrimination of Popular GMF Models}

\citet{Sun08} presented a thorough investigation of  three models for the Galactic magnetic field
which represented a culmination of a variety of inputs from earlier popular models. The three models 
they present are a BSS model and two ASS models: one with magnetic
field reversals following the spiral arms of the Galaxy (ASS+ARM), and the
other with reversals in rings of constant radius (ASS+RING). 
Part of the efforts by \citet{Sun08} included an exploration of how well these models fit 
pulsar RMs \citep{Han99} and  previously published
extragalactic RMs from the

\hspace{40mm}

\hspace{40mm}

\hspace{40mm}

\hspace{40mm}

\hspace{40mm}

\hspace{40mm}

\hspace{40mm}

\hspace{40mm}

\hspace{40mm}

\hspace{40mm}

\hspace{40mm}

\hspace{40mm}

\hspace{40mm}

\hspace{40mm}

\hspace{40mm}

\hspace{40mm}

\hspace{40mm}

\hspace{40mm}

\hspace{40mm}

\hspace{40mm}

\hspace{40mm}

\hspace{40mm}

\hspace{40mm}

\hspace{40mm}

\hspace{40mm}

\hspace{40mm}

\hspace{40mm}

\hspace{40mm}

\noindent   Canadian Galactic Plane Survey \citep{CGPSRMs}, the
Southern Galactic Plane Survey \citep{Brown07}, and a sample of RMs along
sight lines close to the Galactic centre \citep{RoyRMs}.   Using
high-latitude CGPS data \citep[see][]{rae10},  
\citet{Sun08} show an RM  latitude
profile between $100^\circ < \ell < 120^\circ$ (their Figure 12) which clearly
suggests that only the ASS+RING or ASS+ARM models are good contenders, particularly in this
longitude range.  However, as stated by \citet{Ronnie}, filling in the gaps of
 the EGS RM data in the disk is essential to properly discern between the
 models.  With our new data, we are now in a position to contribute to the assessment 
 of these three models. 

\pagebreak

\begin{figure}[htbp]
\plotone{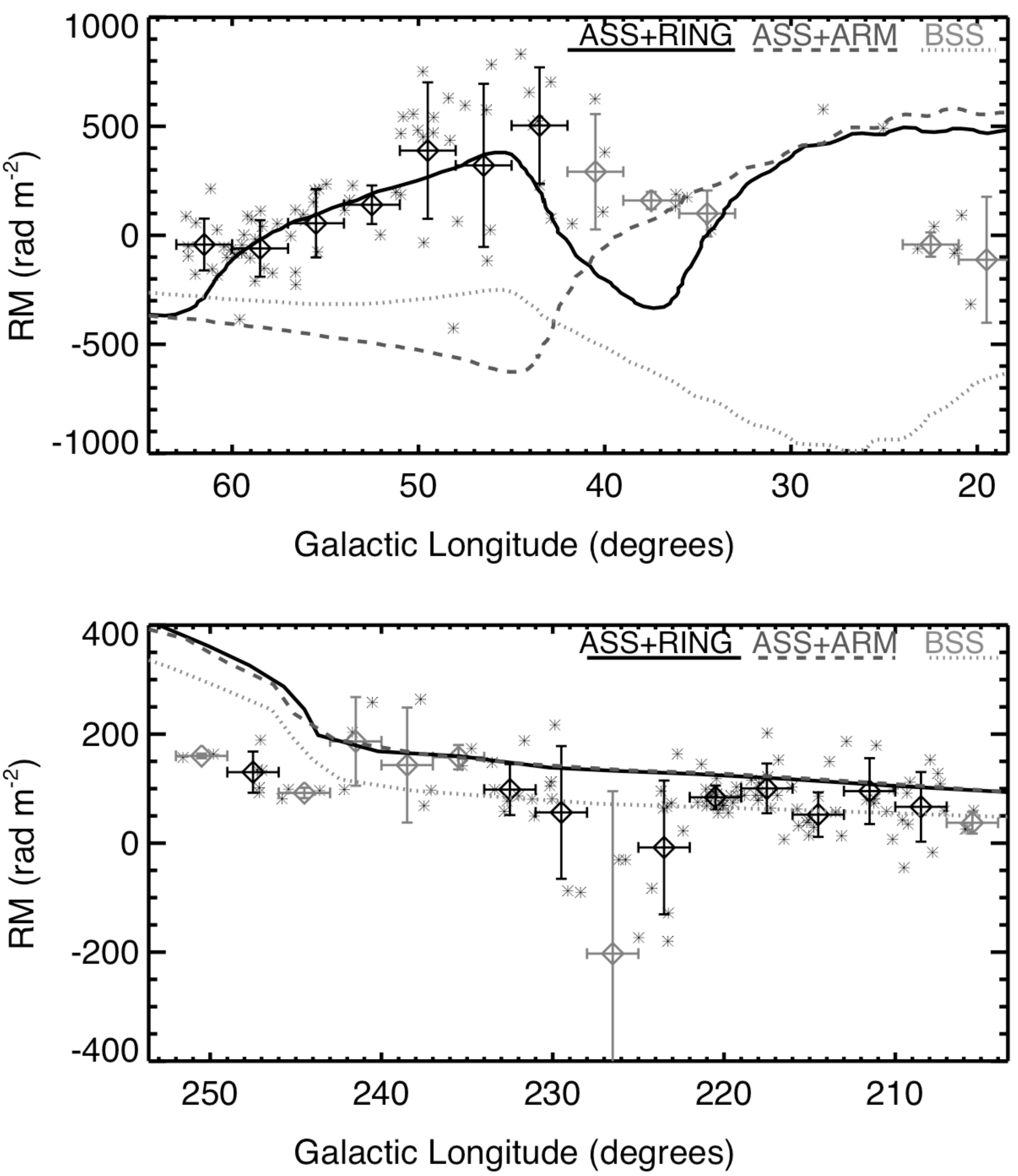}
\caption{Rotation measure versus Galactic longitude for sources in quadrant 1
(upper panel) and quadrant 3 (lower panel).  The diamonds are 3 degree-averaged independent
bins (error bars are the bin width and standard deviation of the mean) of
the individual sources (shown as black asterisks).  Bins shown in black contain at least 5 sources; bins
shown in grey contain between 2 and 4 sources.  The three lines show the
predictions of the models investigated by Sun et al. (2008) taken directly from their Figure 10,
as specified in the upper right corner of each panel.}
\label{fig:RMtrends}
\end{figure}

 Figure \ref{fig:RMtrends}  shows our RMs in the Q1 and Q3 regions seperately,
with the predictions of the models from \citet{Sun08} overlaid.
 As shown, the
 three models have very different predictions in EGS RMs in Q1
 where the reversal's tangent point occurs,
 but,  as could be expected,
 the three
 models are very similar in Q3.
Our data from Q1 is most consistent with the ASS+RING model, 
particularly between $40\degr  < \ell <60\degr$, where
the RMs being generally positive and decreasing with increasing
longitude. The other two models predict negative RMs in this region. 

The ASS+RING model is also more consistent with dynamo models where the
 axisymmetric  solutions of the
 mean-field dynamo equations in a thin disk are of the form
 $\mathbf{B}_u=Q(R)\mathbf{b}(z,R)$, where $Q(R)$ describes the field strength
 along the radius, $R$,  and $\mathbf{b}$  describes the field distribution perpendicular
 to the disk   \citep[with $R$ and $z$ the cylindrical
coordinates;][]{RSS85,PSS93}. A reversal of an axisymmetric field occurs
at $R=R_0$ where $Q(R_0)=0$ (i.e. on a circle).
   We note, however, that none of the
models discussed here fit the observations
as well as
would be desired.  The idea that the field is not as simple as an axisymmetric
or bisymmetric spiral has been discussed extensively by \citet{Men08}.

\section{Multi-sector Model of the Magnetic Field in the Galactic Disk}

 As discussed by \citet{Ronnie}, none of the global models of the large-scale magnetic field  studied to date
adequately reproduce the data across the entire disk.   In the   previous
 section, we determined that the ASS+RING described by \citet{Sun08} fit best for
 our Q1 data.  This  model, however, is quite different than that of  \citet{Brown07} 
 that was quite successful in
 reproducing the data in Q4.  Conversely, the model of \citet{Brown07} does
 not do well in Q1, as demonstrated by \citet{Ronnie}.     
 Furthermore, as the CGPS observations have progressed,
 it has become evident that a simple spiral model in the outer Galaxy is not
 consistent with the data \citep{rae10}.
 We also note that \citet{S05} suggested that the reversal in the Milky Way may be localized in
 a region within  several kiloparsecs near the Sun. If this is the case,
 an entirely different type of
 analysis is required, such as what we propose below.

\begin{figure}[htbp]
\plotone{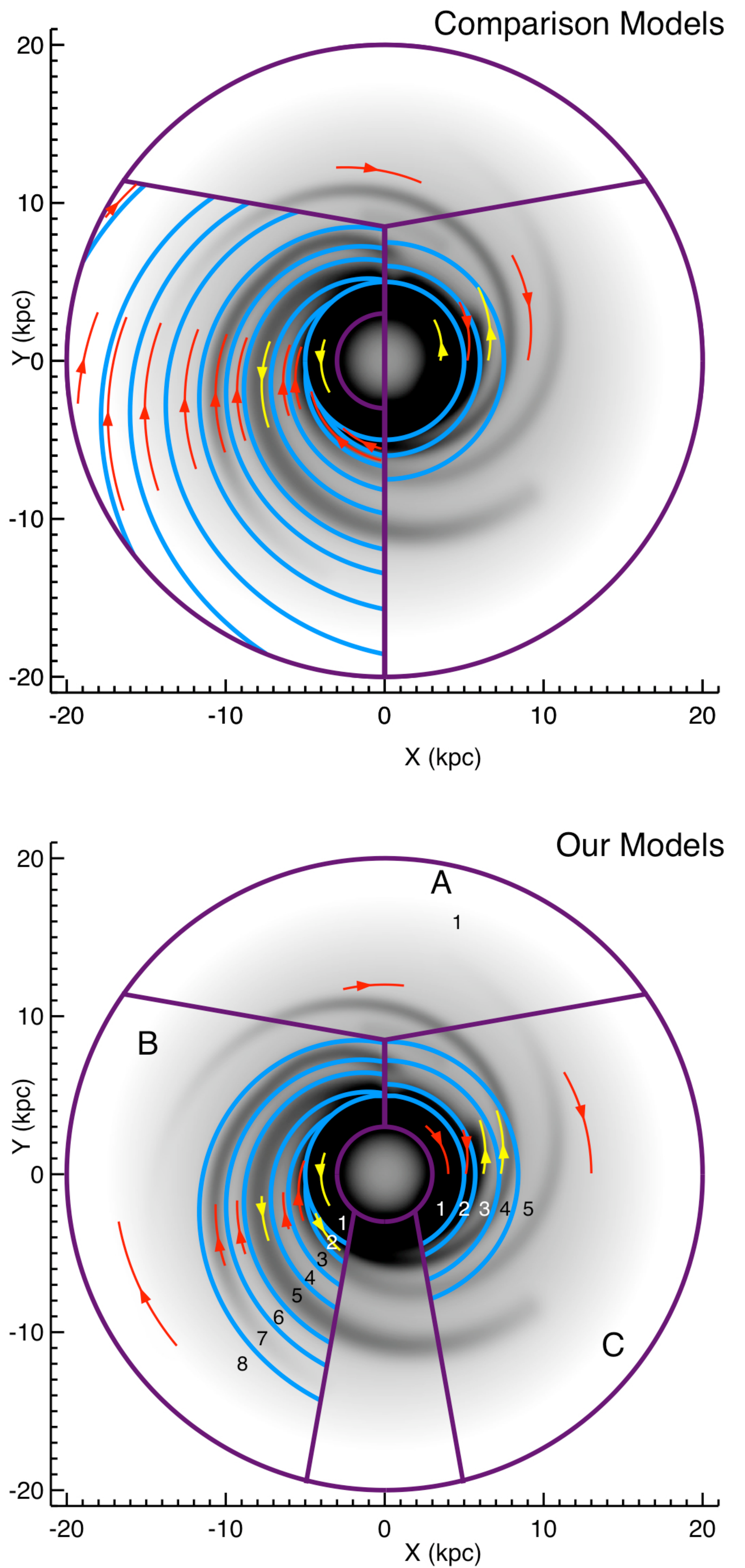}
\caption{Comparison of magnetic field regions and delineation for the 3 sectors investigated in this paper.  Top panel shows
the 3 sectors with previous or comparative model delineations: Region A: Logarithmic spiral with 11.5$^\circ$ magnetic pitch angle; Region B: Model from \citet{Brown07}; Region C: ASS+RING model from \citet{Sun08}. The bottom panel shows the region delineations and magnetic field directions for our  proposed models in the three sectors.
The numbers correspond to the regions listed in Table \ref{table:2}.  We note that the field directions indicated on the lower panel are an output of the model, and were not assigned {\it a priori}.  }
\label{fig:Model_diagrams}
\end{figure}

 While many other
 recent works  have focused on building an empirical model of the large-scale magnetic
 field for the entire Galactic disk \citep[e.g.][]{Weisberg04,Han06, Vallee08} 
 or have looked at only specific regions of the disk \citep[e.g.][]{Brown07, Katgert}, 
   we chose to take a `hybrid' approach.  Our modeling work examines the entire
 Galactic disk, but in 3 separate sectors to see if we can determine any
 common features or structure. We purposely do not apply any `boundary
 matching' conditions between the sectors in order to facilitate independent
 results for each of the three sectors examined.  It was our intention to see
 if there was any commonality amongst the different sectors that could be
 arrived at independently.   With this in mind, we chose the  following three
 sectors, as illustrated in Figure \ref{fig:Model_diagrams}: Sector A isolates
 the outer Galaxy (excluding the local arm) and spans most of quadrants 2 and
 3 (Q2 and Q3); Sector B  spans all of quadrant 4 (Q4) which includes the half of the inner Galaxy 
 with significant lines of sight along the spiral arms; Sector C spans all of quadrant 1 (Q1) which
 includes the half of the inner Galaxy that has lines-of-sight primarily perpendicular
 to the spiral arms.  

 The method we use is described by \citet{Brown07} and in greater detail in
 \citet{joannethesis}.  In summary, we attempt to empirically reproduce the
 observed RMs of both pulsars and EGS using the electron density of
 \citet[][hereafter NE2001]{NE2001}, and various magnetic field models.  We
 acknowledge that this modeling technique is limited in the sense that it
 relies heavily on the validity of the assumed electron density
and magnetic field models.
However, the NE2001 model
reproduces well
the dispersion measures of pulsars  and observed spiral structure at low
latitudes, though it does have limited value at mid to high latitudes
\citep{Gaensler}.  In our case, we are only interested in low latitudes, so
with all other caveats, we have chosen to use this electron density model.
While it is possible to construct magnetic field models using a constant
electron density \citep[e.g.][]{Katgert},  such models put all of the
structure of the RMs into the magnetic field.  Instead, we were interested in
attempting to explore the relationship between the spiral arms and the
magnetic field.

 The magnetic field models investigated here are defined in terms of
several regions with distinct magnetic field configurations (direction and
magnetic pitch angle),
as illustrated
in Figure {\ref{fig:Model_diagrams}}, but the strength and direction of the
field are outputs obtained minimizing the difference between the data and the
model through linear inversion theory \citep{menke}.  By definition within the
model, positive fitted
field strengths correspond to a counter-clockwise field
(as viewed from the North Galactic pole) within that region, while negative
fitted field values correspond to a clockwise
direction.
In addition, since the
goal of our modeling is to explore the large-scale field, we ignore the
small-scale clumps and voids of NE2001, and use only the thin, thick, and
spiral arm components.

In addition, we placed the following restrictions on all of the models we investigated. First, the magnetic field
for Galactocentric radii
$R > 20$ kpc or $R<3$
kpc was set to zero. Similarly, the field was assumed to be
zero for
$|z| > 1.5$ kpc.
  In addition, all models have a circular region containing the molecular ring
of the NE2001 electron model (3 kpc
$\le R \le 5$ kpc)
with a circular magnetic field regardless of the geometry being tested in the rest of
the Galaxy.

 Using the best-fit values for the magnetic field regions in each model (given in Table \ref{table:2}),  we
are then able to determine
the model
RMs for each of the observed source
locations. To assess the quality of fit of each model, we calculated the 
root mean square of the residuals in RMs,
\begin{equation}
\langle (\Delta {\rm{RM}})^2 \rangle ^{1/2} = \langle (\rm{RM_{observed} - RM_{model}})^2 \rangle^{1/2},
\end{equation}
where the goal was to minimize this value with the modeling.
Due to the intrinsic scatter of the EGS RMs and the high angular
density of EGS available to us, we smoothed the observed and
modeled EGS RMs, as described below, before calculating the  $\langle (\Delta {\rm{RM}})^2 \rangle ^{1/2} $
value.

We used EGS RMs from the data presented here, as well as the SGPS and CGPS
data sets.   We treated the EGSs as being located at the edge of the model
($R=20$ kpc).  Since we did not  wish to investigate the complex nature of the
field likely to be found near the Galactic center,
we did not use
any EGS RM sources within $\pm 10^\circ$ Galactic center and consequently did
not use any of the RMs determined by \citet{RoyRMs}. We also did not consider
any vertical structure in our model. Therefore, we removed any sources with a
calculated height
$|z|>1.5$ kpc.   With these criteria, we were left with
184 of the 194 RMs described in section 2, 142 of the 148 RMs from the
SGPS \citep{Brown07} and 1020 sources from the CGPS,
a total of 1346 EGS
sources.

We used 557 pulsar RMs from the following sources:
\citet{Noutsos}, \citet{Han06}, \citet{Weisberg04}, \citet{mitra}, \citet{Han99}
and \citet{pulsars}  where pulsars were selected with
$|z| < 1.5 $ kpc.  For
self-consistency,
we used distances to the pulsars
predicted by the NE2001 model.

The EGS and pulsar RMs were then split by sector
as described in the following sections.
The primary objective of our
modeling
is to produce the best-fitting empirical model with the fewest
parameters.  To that end, we decided to explore the outer Galaxy first, since
it can be
expected to be the simplest in nature.

\subsection{Model Sector A: The Outer Galaxy (Q2 and Q3)}

We define Sector A to be
$100^\circ < \ell < 260^\circ$.  In this
region, we have RMs for 88 pulsars, and 847 EGS (21 from SGPS, 108 from the
catalog in this paper, 718 from CGPS).

As demonstrated by earlier CGPS work, RM data in the outer Galaxy holds no
strong evidence for
a large-scale reversal \citep{bt01, joannethesis,
Brown03}.   There is some evidence that suggests that
the field decays as $R^{-1}$,
consistent with the decay in electron density \citep{Brown03}.    Since the relationship
between the large-scale magnetic field and the electron density has not been formally identified, 
 we investigated several different radial profiles for the large-scale
field including $R^{-1}$, $R^{-1/2}$, $\exp{(-R/5\,{\rm{kpc}})}$, and $\exp{(-R/10\, {\rm{kpc}})}$ and
constant field strength.  All profiles produce similar results in the modeling, as there
are not enough pulsar data to effectively constrain the radial profile of the large-scale magnetic
field  in the outer Galaxy.
Therefore,  we
modeled this region as a single magnetic entity with an  $R^{-1}$ decay profile,
 consistent with earlier observational suggestions.  
The question we were interested in addressing was whether a spiral field was
more or less appropriate than a predominantly azimuthal field.  Recent work by
\citet{rae10} provides evidence for a very small pitch angle in the outer Galaxy, by 
determining the value for the `RM null point' in the outer Galaxy (the longitude where the
RMs transition from positive to negative, corresponding to where the magnetic field is, 
on average, perpendicular to the line of sight) as  $ \ell = 179^\circ \pm 1^\circ$. 
We investigated a range of pitch angles and found that those close to 
zero were clearly preferred as indicated by a minimum in $\langle (\Delta {\rm{RM}})^2 \rangle ^{1/2}$ 
at a pitch angle of $0^\circ$ and a 10\% change in $\langle (\Delta {\rm{RM}})^2 \rangle ^{1/2}$  
occurring  at a pitch angle of $4^\circ$.  

In Figure \ref{fig:Q23}, we present the results for a spiral model inclined at 11.5$^\circ$,
consistent with the spiral arms of NE2001, and compare the results with the
best fit for a purely azimuthal model (magnetic pitch angle of $0^\circ$).
Both  models reproduce the gentle trend of the data to change
from positive RM in Q3  to negative RM in Q2.  However, the purely azimuthal model 
minimizes the
$\langle (\Delta {\rm{RM}})^2 \rangle ^{1/2}$ by
30\% better than the logarithmic spiral model.  
Figure \ref{fig:Q23} also serves to illustrate that  the observed `null point' of the CGPS data clearly occurs close to
$180^\circ$ longitude as predicted by the circular model, implying a field with a 
very small or even zero pitch angle compared to the spiral arms.  This is in contrast with the predicted null of 
$\ell = 166^\circ$ predicted by the spiral model with the same pitch as the spiral
arms \citep[as defined by][]{NE2001}.  The slight shift of the
null-point in the data to $\ell > 180^\circ$  is more likely due to small-scale structures
such as supernova remnants dominating the line of sight magnetic field at
these longitudes \citep[e.g.][]{kothes}, rather than a spiral field with
the opposite `handedness' to that of the optical spiral arms.

 \subsection{Model Sector B: SGPS region (Q4)}

 We define Sector B to be  $260^\circ < \ell < 360^\circ$, which is slightly smaller than the area modeled by \citet{Brown07}.
This region contains 292 pulsars and 121 EGS (all from the SGPS).

 For their model, \citet{Brown07} used the available pulsar RMs combined with
new EGS RMs from the SGPS to model the magnetic field within the SGPS region.
 The model had magnetic regions delineated by the spiral arms of NE2001, with
 a magnetic pitch angle  of 11.5$^\circ$ in all regions, except the `molecular ring'
 which is modeled as an azimuthal field.  For
 $R < 3 $ kpc or $R  > 20 $ kpc 
 the field was assumed to be zero, and there was no vertical component assumed for
 the field. Their field
 strength was
 assumed to have a
$R^{-1}$
 dependence,
 which facilitated a model with significant difference in strength between the
 inner and outer Galaxy.  This was needed since some of the regions appeared
 in the modeling region twice -- once in the inner Galaxy, and once in the outer Galaxy -- as 
 a consequence of the spiral geometry (e.g. the Norma arm in their Figure 4).

Given how well the model of \citet{Brown07} agreed with the data, we decided to keep much of this
 model the same. To that end, we merged all separate regions beyond the
 Sagittarius-Carina arm into one magnetic field region, and redefined the
 field in this new outer-Galaxy region to be purely azimuthal, and still
 retained the  $R^{-1}$
  dependence in this region, consistent with \citet{Brown03}.
 However, for the remaining regions in the inner Galaxy, we reverted to a
 constant field strength, as suggested by \citet{Katgert}.

The best-fit magnetic field results for our variation on the
 \citet{Brown07} model are virtually indistinguishable from the original.  As
shown in Figure \ref{fig:Q4},  the $\langle (\Delta {\rm{RM}})^2 \rangle ^{1/2}$ for our
new model is not statistically significantly different from that for the model
of \citet{Brown07}.  However, the model is less complicated as illustrated in
Figure \ref{fig:Model_diagrams}.  It has 8 regions compared to 9 regions in
the original model, and has removed the complexity of the
$R^{-1}$
 dependence in
the inner Galaxy. The reduction in
the number of
parameters while maintaining a good quality
of fit, makes this new model more attractive.

 \subsection{Model Sector C: VLA region  (Q1) }

We define Sector C to be  $0^\circ < \ell < 100^\circ$.  In this region, we have 177 pulsar RMs and 378 EGS RMs (302 from CGPS, 76 from our VLA observations).

In this segment of the Galaxy, we used as our starting position the  ASS+RING
model described by \citet{Sun08}. In particular, we began by  assuming that in
the inner Galaxy, the magnetic region delineations are circular, but the
fields within the regions are spiral. We note a few critical differences between
our model and theirs:  1) we use only the 
smooth components of NE2001 as noted above (Sun et al.  used all components);  2) we use a maximum
scale height of the field as 1.5 kpc (Sun et al. use 1.0 kpc);  3) we have five separate regions (Sun et al. have four);  4)  we use a magnetic pitch angle within the
individual regions of 11.5$^\circ$, except for the
innermost
region (1C as labeled on Figure \ref{fig:Model_diagrams}) and
the outermost
region (5C), which we define to be azimuthal to be consistent
with our work in sectors A and B  (Sun et al. used 12$^\circ$ in all regions).

The  first region boundary of our model is located at
$R=5.0$
 kpc to correspond to
the molecular ring of NE2001.  The remaining
boundaries were optimized using the $\langle (\Delta {\rm{RM}})^2 \rangle ^{1/2}$ value
as the quality-of-fit measure.  The optimized boundary locations were found to
be
$R=5.8$ kpc, $R=7.2$ kpc, and $R=8.4$ kpc.

For this model, our best fit magnetic field model is predominantly clockwise with a  reversed (clockwise) region in the inner Galaxy, as shown in Figure \ref{fig:Model_diagrams}.
Figure \ref{fig:Q1} shows a direct comparison to the data and values predicted by the ASS+RING model and our model.  When
the data are smoothed, our model shows a factor of 4 improvement in the $\langle (\Delta {\rm{RM}})^2 \rangle ^{1/2}$ over the ASS+RING model.

\subsection{Combining the Sectors}

When we consider our three sectors together, as shown in Figure
\ref{fig:Bmap}, a picture emerges of a predominantly clockwise Galactic
magnetic field with what could be interpreted as single reversed
(counter-clockwise) region spiraling out from the Galactic center, as illustrated
in Figure \ref{fig:Bsketch}.
According to our analysis,
the field
in the inner Galaxy
has a spiral shape (with a pitch angle estimated here as $11.5^\circ$) and is
generally aligned with the spiral arms while in
the outer Galaxy
it is (almost)
azimuthal.   This is consistent with the observations of \citet{Kronberg} who
suggested explicitly that the Galaxy is a mix of an axisymmetric field in the
outer Galaxy and a bisymmetric field in the inner Galaxy. 
The opposite signs of the two molecular rings may be 
suggestive of a bi-symmetric field originating from the Galactic bar.  
More data and perhaps a new electron density model containing the bar would be necessary
in order to properly investigate this region.

We also note that this `spiraling-out'  reversed region could extend
into Q1 at larger Galactic radii, but without any data from pulsars located on the far
side of the Galaxy in this region, determining the existence of such a region
is not possible.
Our model may be considered as something of a zeroth-order approximation; it was constructed in
a piece-wise manner, yet there is some consistency across the whole Galactic disk.
The discontinuities that occur at the boundaries are a consequence of this and indicate that the actual 
field configuration cannot be fully modeled using the simple geometries that we have tested.
Exploration
of more complex geometries  will be necessary to improve the boundary matching.  However, if such
a model significantly increases the number of parameters within the model, additional data will certainly
be needed to properly constrain the model.

\pagebreak

\hspace{5mm}

\begin{figure}[htbp]
\begin{minipage}{\textwidth}
\plotone{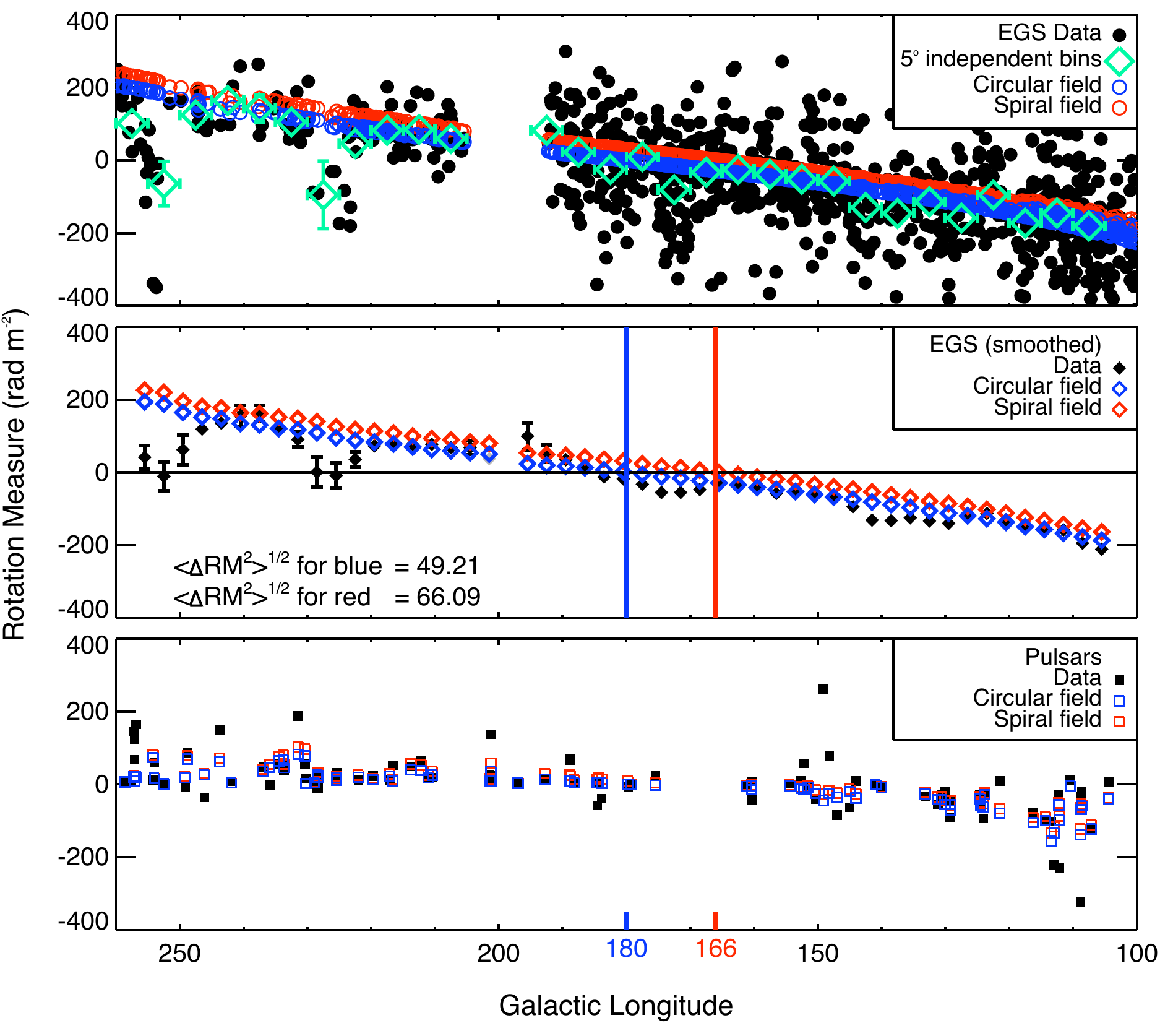}
\caption{Comparison RM versus longitude graphs for observations and models in  Sector A.  The top panel shows the individual EGS RMs from observations (filled black circles) and those data averaged into  5 degree independent bins (green diamonds). The modeled EGS RMs for the best-fit model based on our circular field  are shown as open blue circles,  and  those for a logarithmic spiral with a pitch angle of 11.5$^\circ$  are shown as open red circles.  The middle panel shows the same  data as in top panel, except that the corresponding data have been  box-car averaged over 9$^\circ$ in longitude with a step size of 3$^\circ$. Bin symbols in the middle panel shaded grey indicate bins with fewer than 5 sources.  The mean [median] number of sources per bin is 47 [56]. The green and blue vertical lines show the corresponding RM null points for the two different modeled RM data sets.   The bottom panel shows the RMs  of pulsars for observations and the same models.}
\label{fig:Q23}
\end{minipage}
\end{figure}

\pagebreak

\hspace{40mm}

\pagebreak

\hspace{5mm}

\begin{figure}[htbp]
\begin{minipage}\textwidth
\plotone{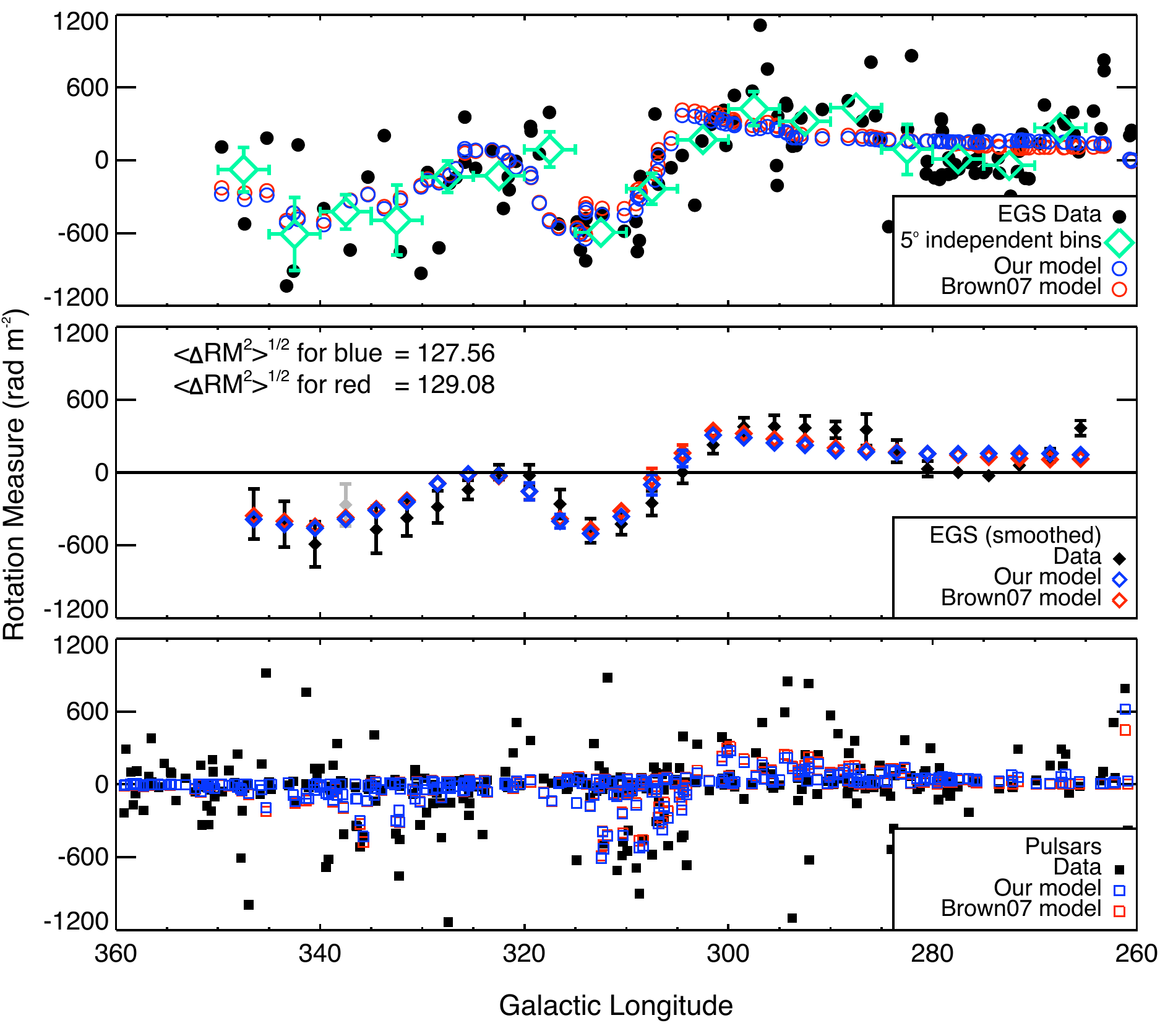}
\caption{Comparison RM versus longitude graphs for observations and models in  Sector B.  The panels are the same as in Figure \ref{fig:Q23} with the comparative model (in red)  being that from \citet{Brown07} and our model (in blue) as described in the text. The mean [median] number of sources per bin in the middle panel is 12 [12]. }
\label{fig:Q4}
\end{minipage}
\end{figure}

\pagebreak

\hspace{40mm}

\pagebreak

\hspace{5mm}

\begin{figure}[htbp]
\begin{minipage}\textwidth
\plotone{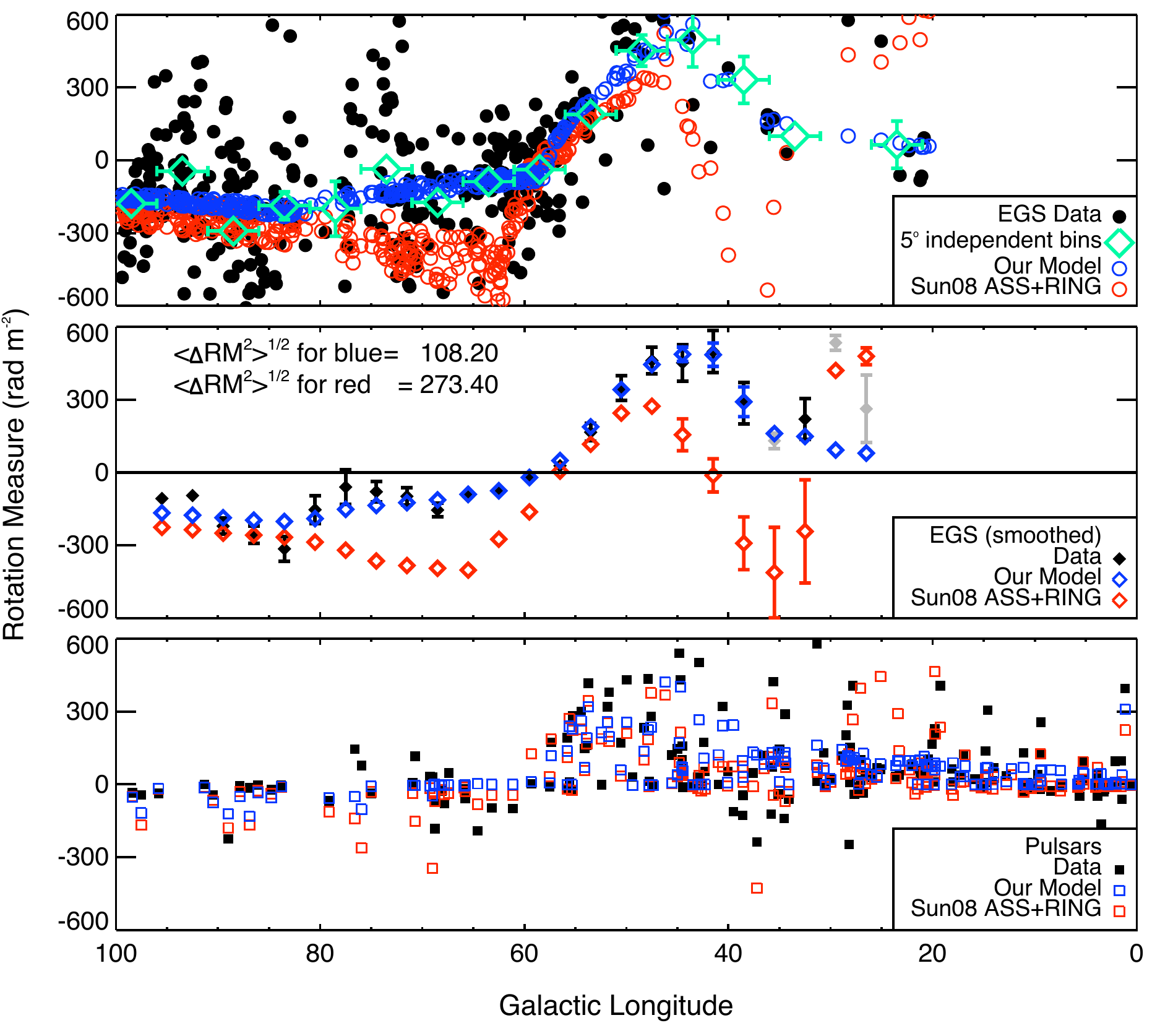}
\caption{Comparison RM versus longitude graphs for observations and models in  Sector C.  The panels are the same as in Figure \ref{fig:Q23} with the comparative model (in red)  being that from \citet{Sun08} and our model (in blue) as described in the text. The mean [median] number of sources per bin in the middle panel is 44 [49]. }
\label{fig:Q1}
\end{minipage}
\end{figure}

\pagebreak

\hspace{40mm}

\pagebreak

\pagebreak

\hspace{5mm}

\begin{figure}[htbp]
\begin{minipage}\textwidth
\plotone{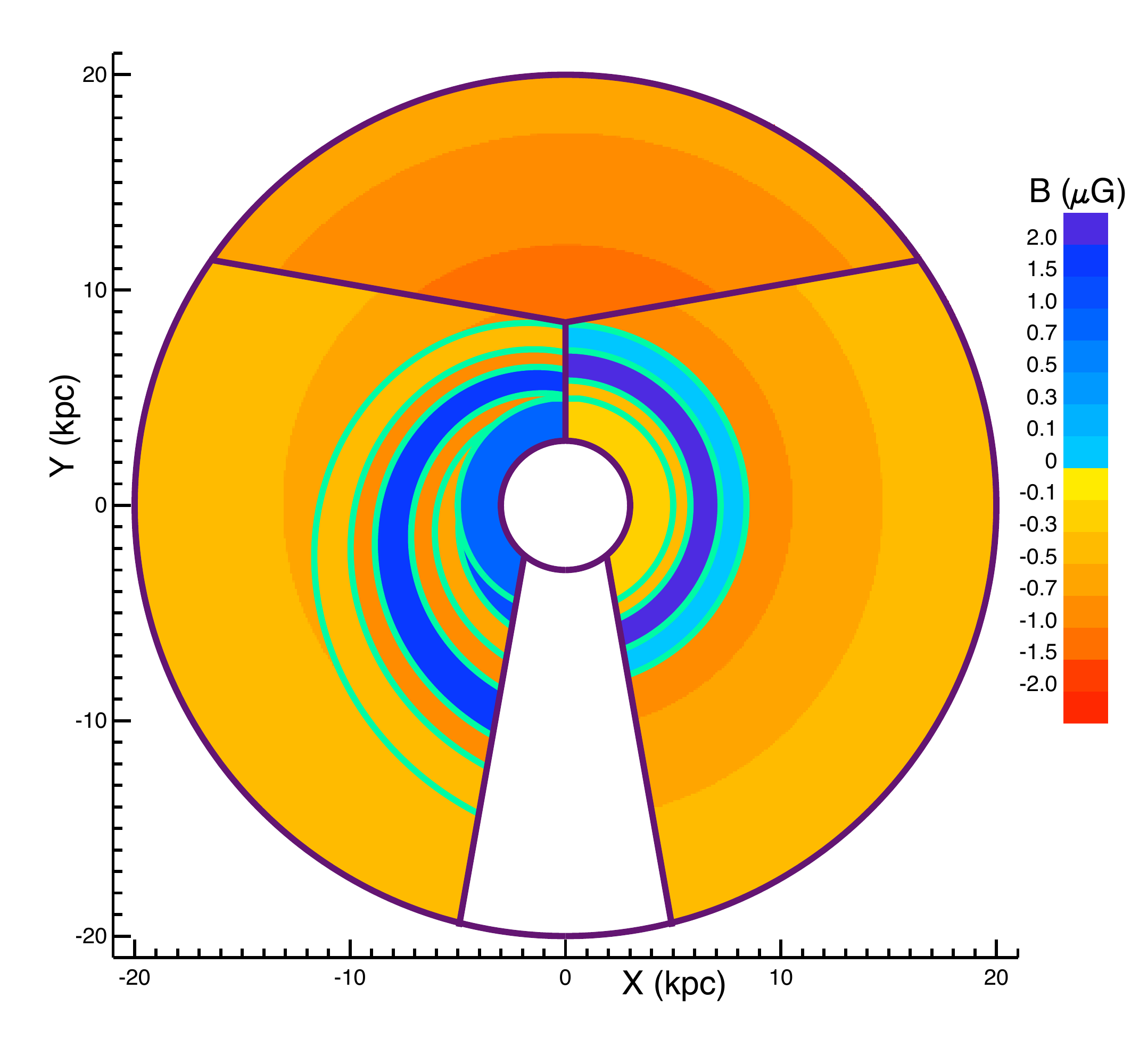}
\caption{Best-fit magnetic field strengths for each of the regions shown in the lower panel of Figure \ref{fig:Model_diagrams}.  Shades of orange/red represent clockwise field, while shades of blue represent a counter-clockwise field. }
\label{fig:Bmap}
\end{minipage}
\end{figure}

\pagebreak

\hspace{40mm}

\pagebreak

\begin{figure}[htbp]
\plotone{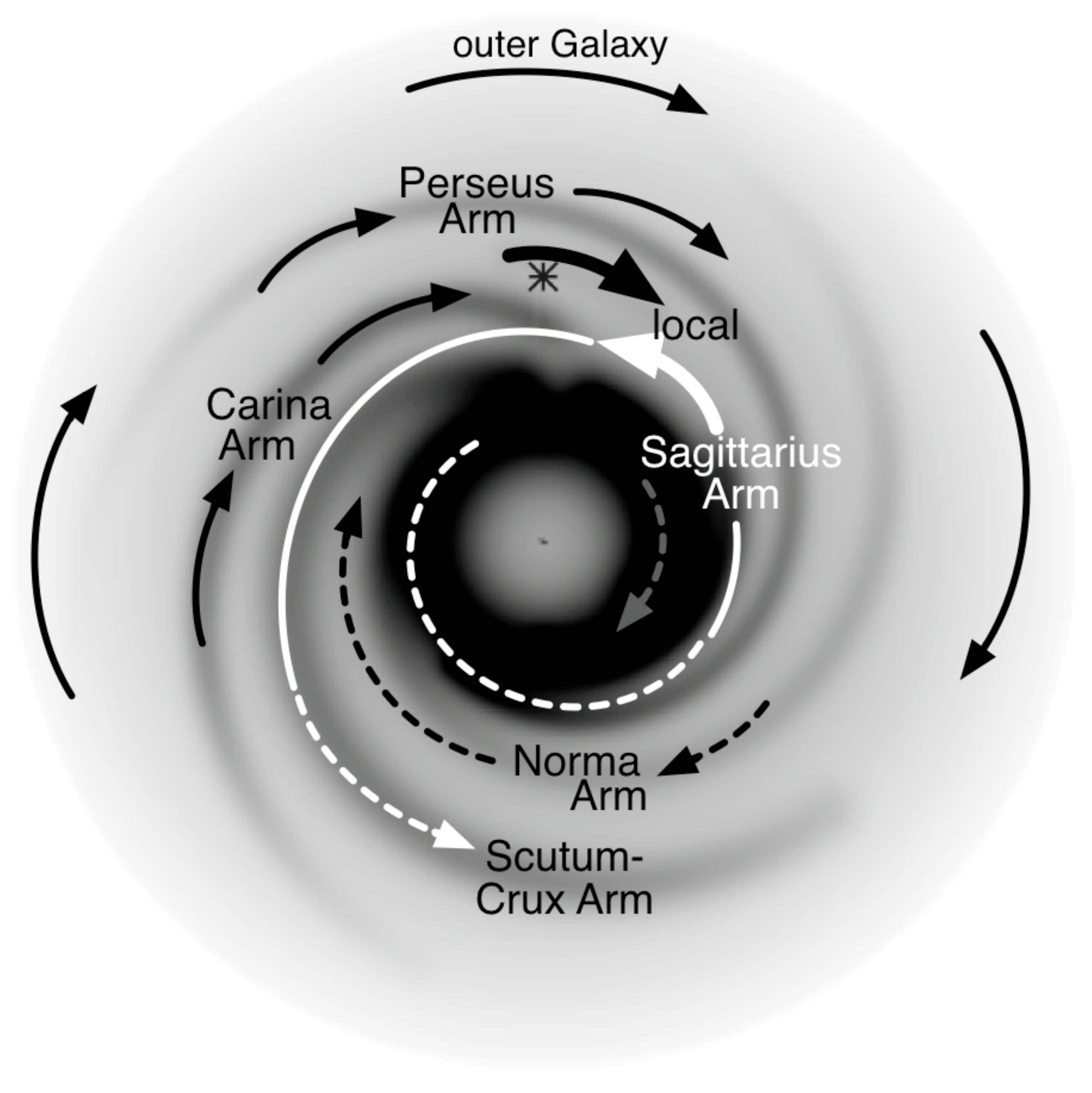}
\caption{A sketch of the magnetic field in the disk of the Galaxy based on this work. The bold arrows in the local arm and Q1 of the Sagittarius-Carina arm shows the only generally accepted location of the large-scale reversal in Q1 \citep[see discussion in][]{Brown11}.  The remaining arrows show the field directions as
concluded from this study.
The dashed arrows are less certain due to the paucity of data available in these regions. }
\label{fig:Bsketch}
\end{figure}

\section{Summary and Discussion}

We have processed a set of VLA observations and produced a catalog of 194
rotation measures of extragalactic sources  in the Galactic disk, filling in
critical gaps in rotation measure coverage  between the Canadian Galactic
Plane Survey and the Southern Galactic Plane Survey.  Using these data, we
conclude that of the three popular models investigated by \citet{Sun08}, the most
consistent with our new data is the ASS+RING model. 

We propose our own model, stemming from a new modeling strategy that studies
the disk field in three different sectors. The division of sectors is  roughly
between the outer Galaxy (quadrants 2 and 3), quadrant 1 and quadrant 4.  Our modeling suggests
that the inner Galaxy has a spiral magnetic field that is aligned with the spiral arms, while
the outer Galaxy is dominated by an almost purely azimuthal field. 
This is consistent with a significant decrease of the magnetic pitch
angle with the galactocentric radius, to small (almost zero) values beyond the Solar
orbit. Such a decrease is also seen in  external spiral galaxies \citep[Fig.~8
in][]{Beck96}.  For example,  the pitch angle in M31 decreases from $-19\pm3^\circ$ near the 
galactic center  to $-8\pm3^\circ$ at
$r=12$--14~kpc \citep{FBBS04}.

Our model also indicates that the magnetic field in the Galaxy
is predominantly clockwise, with a single reversed region that appears to
spiral out from the center of the Galaxy.  This is similar to the ASS+ARM model
described by \citet{Sun08}, except that the pitch angle varies with radius in our model. 
In some sense, our model provides a `unification' of the two axisymmetric spiral
models discussed by \citet{Sun08}.

The origin of magnetic reversals remains poorly understood. An obvious possibility
to explain them is a bisymmetric magnetic field 
\citep[perhaps of primordial origin; see][and references therein]{SFW86}. A bisymmetric magnetic structure has
reversals between spiral-shaped regions, i.e., both in radius and azimuth.
However, it is now believed that bisymmetric magnetic fields are rare in
spiral galaxies, and that galactic magnetic fields are maintained by some form
of dynamo action \citep{Beck96}. Dynamo mechanisms generally favor
axisymmetric magnetic structures, with non-axisymmetric features resulting
from secondary effects (such as the spiral pattern and/or overall
galactic asymmetry). Our results indicate that the regular magnetic field in
the outer part of the Milky Way is predominantly axisymmetric. \citet{RSS85}
suggested that radial reversals of an axisymmetric magnetic field can be
maintained, for periods comparable to the galactic lifetime, provided the
initial (seed) magnetic field had such reversals, for example if the seed
field was random (resulting, e.g., from the fluctuation dynamo action).
\citet{RSS85} confirmed that a few reversals can persist in the Milky Way if
the half-thickness of the ionized layer is within the range 350--1500 pc,
whereas this range is much narrower in the case of M31, 350--450 pc (these
estimates can be model-dependent). This seems to explain the presence of at
least one reversal in the Milky Way and their absence in M31. Asymptotic
analysis of the mean-field galactic dynamo equations with $\alpha$-quenching
\citep{BSS94} shows that the radial reversals can be persistent at those
galactocentric radii $r$ where
\[
r^2\gamma(r)\left(\frac{1}{r}+2\frac{B_0'(r)}{B_0(r)}\right)+\frac12 r^2\gamma'(r)=0,
\]
where $\gamma(r)$ is the local growth rate of the regular magnetic field due
to the dynamo action, $B_0(r)$ is its local saturation strength (presumably
corresponding to the energy equipartition with the turbulent energy), and prime
denotes derivative with respect to radius \citep[see][for a review]{S05}.
Thus, the occurrence of the reversals is sensitive to rather subtle details of
the galactic dynamo that are poorly known. This severely restricts the
predictive power of the theory and limits the value of numerical results,
which are inevitably obtained with idealized and often heavily parameterized
models.

\begin{figure}[htbp]
\begin{minipage}\textwidth
\plotone{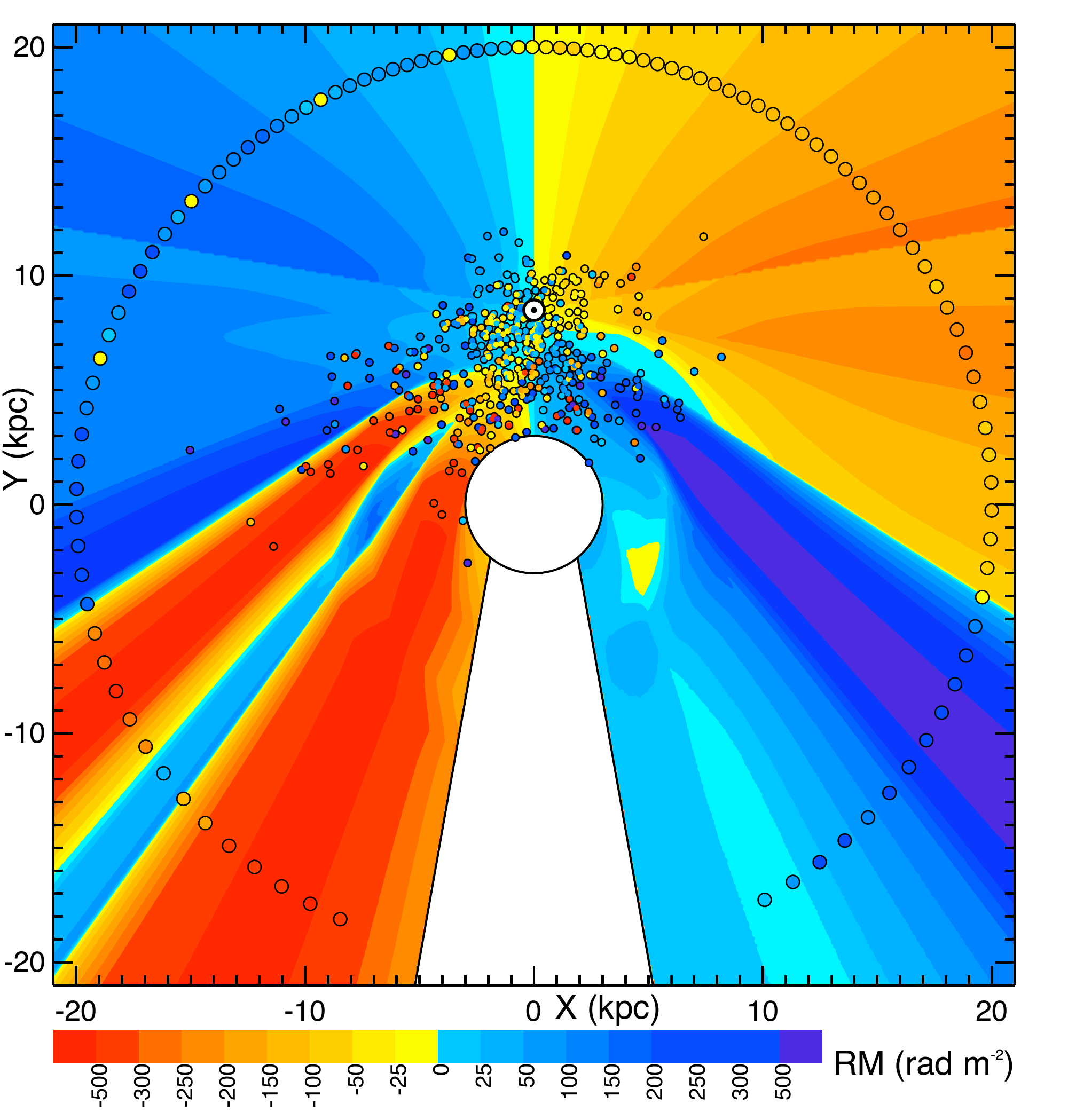}
\caption{Predicted rotation measures for a slice of the Galaxy at  z=0 using our 3 sector model as illustrated in the lower panel of Figure \ref{fig:Model_diagrams}, and Figure \ref{fig:Bmap}.  The smaller circles represent RMs for individual pulsars, while the large circles represent RMs for boxcar-averged EGS sources over 9$^\circ$ longitude with a step size of 3$^\circ$.  The EGS are placed at the edge of our model (R = 20 kpc).  The central black circle is the inner edge of the molecular ring (R = 3 kpc). }
\label{fig:RMmap}
\end{minipage}
\end{figure}

As another way to visually examine our modeling efforts,  we have combined our 3 magnetic sectors with NE2001
to produce a rotation measure map at
$z=0$, as shown in Figure \ref{fig:RMmap}. 
The small circles along the
interior show the pulsar RMs, and the larger circles around the outside (at
$R=20$ kpc) are the smoothed ($9^\circ$ bin widths, $3^\circ$ steps between
bins) EGS RMs, corresponding to the middle plot in Figures \ref{fig:Q23}
through \ref{fig:Q1}.
While certainly there are places where the data and the model disagree in Figure \ref{fig:RMmap} (likely
due to small scale fluctuations in the data that have not been accounted for
and perhaps also due to the limitations of the model),
overall the data appears to be fitted
quite well.  Were it not for the black circles
on the data points, many of these points would be virtually indistinguishable
from the background model.

We expect that significant improvements on this
model, using the same technique and the present edition of the electron
density model, will be difficult to accomplish for several reasons.  First,
the electron density model includes very little small scale structure beyond the
local regions.   Second, the reliability of distances to the pulsars remains
questionable; small shifts in the assumed position of the pulsars will
influence the results of the best fit.  Fortunately, the EGS are simply
assumed to be located at the edge of the Galaxy in their identified  $\ell, b$
direction, making their `position' reliable.  Therefore, they can  provide a
stable base to assess the model.  Finally, as is always the case with
modeling, more data would vastly improve the model.  For example, pulsars on
the far side of the Galactic center would provide much needed constraints 

\pagebreak

\hspace{5mm}

\pagebreak

\hspace{20mm}

\pagebreak

\noindent  in
the area, and as discussed above, the EGS source density is 
considerably lower
in this part of the inner Galaxy as well. 
However, such data are unlikely to
be available
until higher sensitivity instruments, such as ASKAP or the SKA
come online.
While more work and more observations are needed to confirm our findings,  we believe that our approach to
empirical modeling yields a more reasonable model than attempting to model the entire Galaxy with the current status of
both the electron density and available data.

\acknowledgments

We thank the referee for many insightful comments and suggestions that have significantly improved this 
manuscript. 
We also thank Rick Perley and the other staff at the NRAO for their assistance in collecting and processing the data.
This work was supported in part by a grant to J.C.B. from the Natural Sciences and Engineering Research Council of Canada.
B.M.G. acknowledges the support of
an Australian Research Council Federation Fellowship through grant
FF0561298.
The National Radio Astronomy Observatory is a facility of the National Science Foundation operated under cooperative agreement by Associated Universities, Inc.


\pagebreak

\newpage


%
%

\begin{deluxetable}{rrrrrrrrr}
\tablewidth{0 pt}
\tabletypesize{\small}
\tablecolumns{9}
\tablecaption{Rotation Measures from New VLA Observations}
\tablehead{\colhead{$l$} &
\colhead{$b$} & \colhead{$\alpha$} & \colhead{$\delta$} & \colhead{Stokes I$^{\dagger}$} & \colhead{PI$^{\dagger}$} &
\colhead{RM$^{\ddagger}$} &\colhead{RM synthesis} & \colhead{Taylor
RM$^{\ast}$}  \\
\colhead{[\degr]} & \colhead{[\degr]} & \colhead{[h m s]}  & \colhead{[\degr ~  \arcmin ~ \arcsec]} & \colhead{ [mJy]} &
\colhead{[mJy]} & \colhead{[rad m$^{-2}$]} & \colhead{[rad
m$^{-2}$]} & \colhead{[rad m$^{-2}$]}}
\startdata
 20.36 & -1.67 &  18 34            20.9 &  -11 56            37 &  116 &    6 & -317$\pm$ \phantom{0}9 & -318$\pm$   \phantom{0}8 &      --                    \\
 20.84 & -1.66 &  18 35  \phantom{0}6.2 &  -11 30            32 &   45 &    3 &   92$\pm$           11 &   91$\pm$             11 &      --                    \\
 21.08 & -1.60 &  18 35            19.6 &  -11 15            59 &  299 &    9 &  -66$\pm$           12 &  -65$\pm$             13 &   -23.4$\pm$           17.6\\
 21.22 & -2.26 &  18 37            59.8 &  -11 26            27 &  151 &   16 &  -83$\pm$ \phantom{0}4 &  -82$\pm$   \phantom{0}4 &   -75.7$\pm$ \phantom{0}6.0\\
 22.32 & +1.73 &  18 25            42.7 &  -08 37            23 &  353 &   10 &   40$\pm$           13 &   40$\pm$             14 &    56.3$\pm$           11.9\\
 23.20 & +2.21 &  18 25            37.6 &  -07 37            29 &  307 &    9 &  -62$\pm$           13 &  -61$\pm$             13 &   -81.1$\pm$           13.7\\
 25.04 & -2.65 &  18 46            30.0 &  -08 13            33 &  144 &    8 &  491$\pm$ \phantom{0}8 &  492$\pm$   \phantom{0}8 &      --                    \\
 28.27 & -2.24 &  18 50            54.2 &  -05 09            42 &  214 &    8 &  577$\pm$           10 &  575$\pm$             11 &      --                    \\
 34.30 & +2.44 &  18 45            13.6 &  +02 19            54 &   36 &    4 &   25$\pm$ \phantom{0}7 &   27$\pm$   \phantom{0}7 &      --                    \\
 35.57 & +1.40 &  18 51            14.5 &  +02 59            39 &  501 &    9 &  175$\pm$           19 &  175$\pm$             21 &    34.6$\pm$           14.1\\
 36.18 & +1.44 &  18 52            13.4 &  +03 32            55 &  222 &   12 &  188$\pm$ \phantom{0}7 &  188$\pm$   \phantom{0}7 &  -458.4$\pm$ \phantom{0}8.7\\
 36.20 & -0.35 &  18 58            37.8 &  +02 45            18 &   95 &    4 &  132$\pm$           11 &  133$\pm$             11 &      --                    \\
 40.01 & +1.02 &  19 00            43.4 &  +06 45            46 &  151 &   12 &  380$\pm$ \phantom{0}5 &  380$\pm$   \phantom{0}6 &  -252.3$\pm$           13.4\\
 40.10 & -3.61 &  19 17            25.5 &  +04 42            36 &  192 &    6 &  107$\pm$           14 &  105$\pm$             13 &      --                    \\
 40.52 & +1.71 &  18 59            12.3 &  +07 32            14 &  140 &    4 &  625$\pm$           14 &  625$\pm$             14 &      --                    \\
 41.75 & -3.27 &  19 19            17.3 &  +06 19            42 &  262 &    3 &   53$\pm$           32 &   55$\pm$             34 &      --                    \\
 42.88 & -3.59 &  19 22            33.9 &  +07 10            48 &  122 &    2 &   78$\pm$           20 &   75$\pm$             22 &      --                    \\
 42.89 & +0.57 &  19 07            41.9 &  +09 07            17 &  662 &   22 &  703$\pm$           12 &  703$\pm$             12 &      --                    \\
 43.49 & -2.30 &  19 19  \phantom{0}6.3 &  +08 19            20 &   82 &    7 &  229$\pm$ \phantom{0}6 &  230$\pm$   \phantom{0}6 &   191.5$\pm$           15.7\\
 43.58 & +3.74 &  18 57            31.2 &  +11 10            53 &  103 &   14 &  525$\pm$ \phantom{0}4 &  524$\pm$   \phantom{0}4 &      --                    \\
 43.84 & -1.44 &  19 16            41.2 &  +09 01            47 &   35 &    5 &  505$\pm$ \phantom{0}6 &  504$\pm$   \phantom{0}7 &      --                    \\
 44.06 & +3.33 &  18 59            52.5 &  +11 25            14 &  112 &    7 &  655$\pm$ \phantom{0}7 &  655$\pm$   \phantom{0}8 &      --                    \\
 44.51 & +2.60 &  19 03            23.2 &  +11 29  \phantom{0}5 &  270 &   45 &  831$\pm$ \phantom{0}3 &  831$\pm$   \phantom{0}3 &   141.3$\pm$ \phantom{0}7.4\\
 46.09 & +1.59 &  19 10  \phantom{0}0.4 &  +12 25            24 &  197 &    7 &  783$\pm$           11 &  783$\pm$             11 &      --                    \\
 46.14 & -3.76 &  19 29            22.8 &  +09 59  \phantom{0}1 &  711 &   50 &   24$\pm$ \phantom{0}6 &   24$\pm$   \phantom{0}6 &   686.6$\pm$ \phantom{0}7.0\\
 46.31 & -0.38 &  19 17            33.9 &  +11 42            15 &  122 &    7 & -117$\pm$ \phantom{0}8 & -117$\pm$   \phantom{0}8 &   529.2$\pm$           11.6\\
 46.35 & +3.09 &  19 05  \phantom{0}1.7 &  +13 20            47 &  228 &    9 &  575$\pm$           10 &  577$\pm$             10 &      --                    \\
 47.50 & +2.17 &  19 10            31.9 &  +13 58            10 &  162 &    7 &  596$\pm$ \phantom{0}9 &  597$\pm$   \phantom{0}9 &      --                    \\
 47.91 & -1.78 &  19 25            40.8 &  +12 27            38 &  214 &    7 &   63$\pm$           12 &   63$\pm$             13 &      --                    \\
 48.12 & -3.98 &  19 34  \phantom{0}0.2 &  +11 35            50 &   61 &    2 & -426$\pm$           17 & -426$\pm$             18 &      --                    \\
 48.31 & -1.37 &  19 24            58.7 &  +13 00            33 &  151 &    8 &  435$\pm$ \phantom{0}8 &  436$\pm$   \phantom{0}8 &      --                    \\
 48.39 & +3.62 &  19 06            55.4 &  +15 23            42 &  131 &   16 &  629$\pm$ \phantom{0}4 &  629$\pm$   \phantom{0}4 &   -37.5$\pm$           10.1\\
 49.18 & +1.36 &  19 16            44.2 &  +15 03            49 &  425 &   16 &  541$\pm$           10 &  542$\pm$             11 &  -145.5$\pm$ \phantom{0}8.8\\
 49.21 & -0.97 &  19 25            17.3 &  +13 59            19 &  784 &   43 &  470$\pm$ \phantom{0}7 &  470$\pm$   \phantom{0}7 &   442.5$\pm$ \phantom{0}3.6\\
 49.72 & -3.56 &  19 35            38.0 &  +13 11            44 &   29 &    2 &  -34$\pm$           12 &  -34$\pm$             13 &      --                    \\
 49.75 & +3.75 &  19 09  \phantom{0}1.0 &  +16 39            44 &  489 &   19 &  451$\pm$           10 &  451$\pm$             10 &  -235.4$\pm$ \phantom{0}7.9\\
 49.77 & +2.86 &  19 12            19.9 &  +16 16            28 &  464 &   23 &  751$\pm$ \phantom{0}8 &  751$\pm$   \phantom{0}8 &    41.7$\pm$ \phantom{0}8.2\\
 50.04 & +2.06 &  19 15            49.7 &  +16 08            34 &  128 &    7 &  482$\pm$ \phantom{0}7 &  482$\pm$   \phantom{0}8 &      --                    \\
 50.28 & +2.61 &  19 14            14.5 &  +16 36            40 &  276 &   13 &  556$\pm$ \phantom{0}8 &  556$\pm$   \phantom{0}9 &  -123.7$\pm$           15.9\\
 50.82 & +1.19 &  19 20            32.3 &  +16 25            57 &   76 &    7 &  543$\pm$ \phantom{0}6 &  542$\pm$   \phantom{0}6 &      --                    \\
 50.94 & +0.84 &  19 22  \phantom{0}3.8 &  +16 22            43 &  246 &   11 &  466$\pm$ \phantom{0}9 &  466$\pm$   \phantom{0}9 &  -228.1$\pm$           10.7\\
 50.95 & -2.18 &  19 33  \phantom{0}6.4 &  +14 56            24 &  414 &   10 &  186$\pm$           15 &  187$\pm$             16 &   137.0$\pm$           11.8\\
 51.22 & -1.42 &  19 30            52.8 &  +15 32            35 &  709 &   12 &  196$\pm$           22 &  196$\pm$             23 &      --                    \\
 52.04 & -2.67 &  19 37  \phantom{0}4.8 &  +15 39            19 &  256 &   13 &    1$\pm$ \phantom{0}8 &    1$\pm$   \phantom{0}8 &   -37.9$\pm$           11.1\\
 53.54 & +3.13 &  19 18            39.3 &  +19 44  \phantom{0}4 &  753 &  127 &  228$\pm$ \phantom{0}3 &  228$\pm$   \phantom{0}3 &   215.7$\pm$ \phantom{0}0.9\\
 53.70 & +3.30 &  19 18            20.7 &  +19 57            52 &  140 &   13 &  161$\pm$ \phantom{0}5 &  161$\pm$   \phantom{0}5 &   160.5$\pm$           10.6\\
 53.96 & +3.13 &  19 19            29.9 &  +20 06            29 &  258 &   10 &  114$\pm$           10 &  114$\pm$             10 &   129.9$\pm$           13.0\\
 54.95 & +2.30 &  19 24            36.9 &  +20 35            21 &  583 &   37 &  234$\pm$ \phantom{0}6 &  234$\pm$   \phantom{0}6 &   207.5$\pm$ \phantom{0}2.3\\
 55.34 & +3.95 &  19 19  \phantom{0}8.6 &  +21 42            58 &   63 &   12 &  212$\pm$ \phantom{0}3 &  212$\pm$   \phantom{0}3 &   203.1$\pm$ \phantom{0}8.1\\
 55.40 & -2.58 &  19 43            41.5 &  +18 37            20 &  137 &    7 &  -76$\pm$ \phantom{0}8 &  -77$\pm$   \phantom{0}8 &  -120.5$\pm$           17.0\\
 55.56 & +2.26 &  19 25            59.5 &  +21 06            26 & 1650 &   42 &  206$\pm$           15 &  206$\pm$             15 &   217.4$\pm$ \phantom{0}2.9\\
 55.64 & +0.32 &  19 33            25.4 &  +20 14            59 &   78 &    5 &  152$\pm$ \phantom{0}8 &  152$\pm$   \phantom{0}9 &      --                    \\
 56.08 & +0.10 &  19 35            10.4 &  +20 31            54 &  293 &    9 &   67$\pm$           12 &   67$\pm$             13 &      --                    \\
 56.18 & -0.57 &  19 37            54.5 &  +20 17            30 &  114 &    3 &   96$\pm$           15 &   97$\pm$             16 &      --                    \\
 56.59 & -2.39 &  19 45            29.9 &  +19 44            37 &  161 &    9 & -227$\pm$ \phantom{0}8 & -228$\pm$   \phantom{0}8 &  -200.7$\pm$           14.3\\
 56.60 & -1.11 &  19 40            46.6 &  +20 23            51 &   97 &    2 & -170$\pm$           26 & -171$\pm$             27 &      --                    \\
 56.62 & +0.17 &  19 36  \phantom{0}2.4 &  +21 01            47 &  143 &    3 &  117$\pm$           18 &  115$\pm$             19 &      --                    \\
 56.85 & +0.97 &  19 33            31.4 &  +21 37            38 &  134 &    9 &   -4$\pm$ \phantom{0}6 &   -4$\pm$   \phantom{0}7 &      --                    \\
 57.58 & +1.93 &  19 31            24.9 &  +22 43            32 &  568 &   19 &   53$\pm$           11 &   53$\pm$             12 &    28.4$\pm$ \phantom{0}5.4\\
 57.83 & -2.36 &  19 48  \phantom{0}0.7 &  +20 50            26 &  427 &   35 & -174$\pm$ \phantom{0}5 & -174$\pm$   \phantom{0}5 &  -206.9$\pm$ \phantom{0}2.7\\
 58.29 & -2.59 &  19 49            51.7 &  +21 06            56 &  172 &    6 & -152$\pm$           12 & -152$\pm$             13 &      --                    \\
 58.38 & +2.26 &  19 31            49.2 &  +23 35  \phantom{0}2 &   42 &    3 &   41$\pm$           10 &   42$\pm$             10 &      --                    \\
 58.48 & +2.62 &  19 30            37.9 &  +23 50            45 &  290 &    9 &  112$\pm$           12 &  112$\pm$             12 &    74.2$\pm$ \phantom{0}8.6\\
 58.77 & -0.73 &  19 43            57.6 &  +22 27            42 &  133 &    4 &  -17$\pm$           12 &  -15$\pm$             13 &      --                    \\
 58.78 & -2.40 &  19 50            14.6 &  +21 38  \phantom{0}0 &  127 &   16 & -210$\pm$ \phantom{0}4 & -210$\pm$   \phantom{0}4 &  -190.1$\pm$ \phantom{0}5.9\\
 58.97 & +0.20 &  19 40            50.3 &  +23 06  \phantom{0}5 &  265 &   12 &   73$\pm$ \phantom{0}8 &   73$\pm$   \phantom{0}9 &    36.9$\pm$ \phantom{0}6.1\\
 59.03 & +1.85 &  19 34            44.4 &  +23 57            50 &   84 &    3 & -105$\pm$           14 & -110$\pm$             15 &      --                    \\
 59.17 & +2.82 &  19 31            18.6 &  +24 33  \phantom{0}2 &   45 &    2 &   89$\pm$           12 &   90$\pm$             13 &      --                    \\
 59.37 & +3.21 &  19 30            15.4 &  +24 54            29 &  131 &    9 &    2$\pm$ \phantom{0}7 &    1$\pm$   \phantom{0}7 &   -24.0$\pm$           10.3\\
 59.47 & +3.57 &  19 29  \phantom{0}2.2 &  +25 10            35 &   40 &    2 &  -44$\pm$           14 &  -46$\pm$             15 &      --                    \\
 59.48 & +3.42 &  19 29            39.8 &  +25 06            26 &  174 &    9 &  -83$\pm$ \phantom{0}8 &  -83$\pm$   \phantom{0}8 &   -91.6$\pm$           12.2\\
 59.48 & +1.95 &  19 35            20.3 &  +24 24            15 &   17 &    3 &  -79$\pm$ \phantom{0}9 &  -80$\pm$   \phantom{0}9 &      --                    \\
 59.60 & -0.07 &  19 43            18.3 &  +23 30            32 &  171 &    9 & -387$\pm$ \phantom{0}8 & -387$\pm$   \phantom{0}8 &   272.2$\pm$ \phantom{0}8.6\\
 59.87 & -1.35 &  19 48            41.0 &  +23 06  \phantom{0}8 &  200 &    5 &  -85$\pm$           15 &  -84$\pm$             16 &      --                    \\
\enddata
\end{deluxetable}

\newpage

\addtocounter{table}{-1}
\begin{deluxetable}{rrrrrrrrr}
\tablewidth{0 pt}
\tabletypesize{\small}
\tablecolumns{9}
\tablecaption{-- {\sl{Continued}}}
\tablehead{\colhead{$l$} &
\colhead{$b$} & \colhead{$\alpha$} & \colhead{$\delta$} & \colhead{Stokes I$^{\dagger}$} & \colhead{PI$^{\dagger}$} &
\colhead{RM$^{\ddagger}$} &\colhead{RM synthesis} & \colhead{Taylor
RM$^{\ast}$}  \\
\colhead{[\degr]} & \colhead{[\degr]} & \colhead{[h m s]}  & \colhead{[\degr ~  \arcmin ~ \arcsec]} & \colhead{ [mJy]} &
\colhead{[mJy]} & \colhead{[rad m$^{-2}$]} & \colhead{[rad
m$^{-2}$]} & \colhead{[rad m$^{-2}$]}}
\startdata
60.09 & +3.45 &  19 30            49.3 &  +25 39            12 &  170 &    9 &  -67$\pm$ \phantom{0}7 &  -67$\pm$   \phantom{0}8 &   -73.5$\pm$           11.5\\
 60.29 & +1.89 &  19 37            19.0 &  +25 04            31 &   70 &    8 & -102$\pm$ \phantom{0}5 & -103$\pm$   \phantom{0}5 &  -103.5$\pm$           12.3\\
 60.44 & +1.06 &  19 40            48.5 &  +24 47            48 &  113 &    5 &  -54$\pm$           10 &  -53$\pm$             10 &   -19.3$\pm$           13.2\\
 60.78 & -0.64 &  19 48  \phantom{0}2.0 &  +24 14            42 &  337 &   13 & -184$\pm$           10 & -184$\pm$             11 &  -189.1$\pm$ \phantom{0}6.9\\
 60.81 & +1.59 &  19 39            34.2 &  +25 23            20 &  156 &    8 &   24$\pm$ \phantom{0}8 &   24$\pm$   \phantom{0}9 &     9.2$\pm$           11.9\\
 61.07 & -2.08 &  19 54  \phantom{0}5.8 &  +23 45            41 &  110 &    9 & -154$\pm$ \phantom{0}6 & -155$\pm$   \phantom{0}6 &  -165.6$\pm$           11.0\\
 61.16 & +3.92 &  19 31            14.7 &  +26 49  \phantom{0}6 &  172 &   10 &  214$\pm$ \phantom{0}7 &  214$\pm$   \phantom{0}8 &   215.2$\pm$ \phantom{0}8.8\\
 61.95 & +2.39 &  19 38            58.3 &  +26 46            12 &  892 &   16 &   58$\pm$           21 &   58$\pm$             21 &    42.4$\pm$ \phantom{0}4.1\\
 61.98 & -2.33 &  19 57  \phantom{0}5.5 &  +24 24            36 &  377 &    9 & -179$\pm$           15 & -180$\pm$             16 &   -20.2$\pm$ \phantom{0}7.7\\
 62.36 & -0.96 &  19 52            48.7 &  +25 26            58 & 1093 &   40 &  -96$\pm$           11 &  -96$\pm$             11 &  -106.3$\pm$ \phantom{0}2.9\\
 62.36 & -1.77 &  19 55            52.2 &  +25 01            47 &  494 &   36 &  -53$\pm$ \phantom{0}5 &  -53$\pm$   \phantom{0}6 &   -65.5$\pm$ \phantom{0}2.9\\
 62.48 & +1.42 &  19 43            56.4 &  +26 44            56 &   88 &    5 &   86$\pm$ \phantom{0}8 &   86$\pm$   \phantom{0}8 &      --                    \\
205.43 & -3.75 &  06 24            18.8 &  +04 57  \phantom{0}1 &  371 &   13 &   60$\pm$           10 &   61$\pm$             11 &    57.0$\pm$ \phantom{0}6.3\\
205.81 & +4.91 &  06 56  \phantom{0}1.0 &  +08 34  \phantom{0}7 &  503 &   26 &   26$\pm$ \phantom{0}8 &   26$\pm$   \phantom{0}8 &    20.1$\pm$ \phantom{0}4.2\\
205.92 & +3.39 &  06 50            45.9 &  +07 47  \phantom{0}9 &  357 &    9 &   26$\pm$           15 &   26$\pm$             16 &    -6.8$\pm$ \phantom{0}5.7\\
207.14 & -1.33 &  06 36  \phantom{0}5.7 &  +04 32            40 & 1001 &   54 &   59$\pm$ \phantom{0}7 &   59$\pm$   \phantom{0}7 &    50.2$\pm$ \phantom{0}1.5\\
207.37 & +3.62 &  06 54            12.7 &  +06 35            39 &  111 &    7 &  109$\pm$ \phantom{0}7 &  110$\pm$   \phantom{0}8 &      --                    \\
207.52 & +3.25 &  06 53  \phantom{0}9.1 &  +06 18            10 &  589 &   35 &  128$\pm$ \phantom{0}7 &  128$\pm$   \phantom{0}7 &   107.2$\pm$ \phantom{0}2.9\\
207.82 & -4.39 &  06 26            28.0 &  +02 31            51 &  480 &   21 &  -17$\pm$ \phantom{0}9 &  -17$\pm$   \phantom{0}9 &    -0.4$\pm$ \phantom{0}4.5\\
207.97 & +2.14 &  06 50  \phantom{0}0.7 &  +05 23            59 &   92 &    8 &  153$\pm$ \phantom{0}5 &  153$\pm$   \phantom{0}6 &   156.2$\pm$           12.0\\
209.15 & -2.62 &  06 35            10.2 &  +02 10  \phantom{0}9 &  144 &    8 &  112$\pm$ \phantom{0}7 &  111$\pm$   \phantom{0}8 &   124.5$\pm$ \phantom{0}7.7\\
209.25 & -4.63 &  06 28            12.6 &  +01 09            26 &  962 &   30 &   34$\pm$           12 &   35$\pm$             12 &    38.9$\pm$ \phantom{0}3.3\\
209.31 & +4.81 &  07 02  \phantom{0}1.7 &  +05 24            24 &  157 &    4 &   92$\pm$           14 &   94$\pm$             15 &    28.2$\pm$           12.4\\
209.50 & +1.10 &  06 49  \phantom{0}4.6 &  +03 33            51 &  113 &   10 &  -45$\pm$ \phantom{0}5 &  -45$\pm$   \phantom{0}5 &   -58.1$\pm$ \phantom{0}8.1\\
209.59 & +2.38 &  06 53            49.9 &  +04 03            47 &  358 &   14 &   42$\pm$           10 &   42$\pm$             10 &    39.6$\pm$ \phantom{0}7.0\\
210.12 & -2.62 &  06 36            56.6 &  +01 18            22 &  419 &   32 &   99$\pm$ \phantom{0}5 &   99$\pm$   \phantom{0}5 &   158.3$\pm$ \phantom{0}8.1\\
210.16 & -4.08 &  06 31            51.5 &  +00 36            14 &  299 &    7 &    7$\pm$           15 &    6$\pm$             16 &    17.2$\pm$           10.1\\
210.51 & +1.11 &  06 50            59.7 &  +02 39            56 &   93 &    9 &   58$\pm$ \phantom{0}5 &   58$\pm$   \phantom{0}6 &    58.5$\pm$           15.3\\
211.12 & -1.56 &  06 42            32.0 &  +00 54            18 &   99 &    9 &  179$\pm$ \phantom{0}5 &  180$\pm$   \phantom{0}6 &   158.5$\pm$           10.7\\
211.18 & +0.85 &  06 51            16.2 &  +01 56            51 &  209 &   10 &   81$\pm$ \phantom{0}9 &   81$\pm$   \phantom{0}9 &    62.2$\pm$           11.8\\
211.26 & +2.41 &  06 56            57.7 &  +02 35            34 &  173 &   10 &   73$\pm$ \phantom{0}7 &   74$\pm$   \phantom{0}8 &    60.1$\pm$ \phantom{0}7.5\\
211.67 & +0.63 &  06 51            21.9 &  +01 24            27 &  298 &    9 &   79$\pm$           13 &   79$\pm$             13 &     4.5$\pm$ \phantom{0}7.9\\
212.85 & -1.78 &  06 44            56.0 &  -00 44            17 &  246 &    7 &  187$\pm$           13 &  187$\pm$             14 &   159.5$\pm$           14.0\\
213.14 & +4.73 &  07 08            38.6 &  +01 58            25 &  126 &    4 &   13$\pm$           13 &   10$\pm$             13 &      --                    \\
213.49 & +3.62 &  07 05            17.6 &  +01 09            30 &  150 &    9 &   57$\pm$ \phantom{0}7 &   57$\pm$   \phantom{0}7 &    72.1$\pm$           12.1\\
213.84 & -3.60 &  06 40            15.6 &  -02 26            51 &  158 &    8 &  150$\pm$ \phantom{0}9 &  151$\pm$   \phantom{0}9 &   173.8$\pm$           10.3\\
214.49 & -1.08 &  06 50            25.7 &  -01 52            22 &  457 &   12 &   86$\pm$           14 &   86$\pm$             15 &   108.3$\pm$ \phantom{0}6.4\\
214.76 & +4.53 &  07 10            53.6 &  +00 26            48 &  557 &   36 &   29$\pm$ \phantom{0}6 &   29$\pm$   \phantom{0}6 &    32.1$\pm$ \phantom{0}2.0\\
215.05 & +2.26 &  07 03            19.1 &  -00 51  \phantom{0}2 &  936 &   23 &   14$\pm$           15 &   14$\pm$             16 &    17.6$\pm$ \phantom{0}0.7\\
215.07 & +3.44 &  07 07            33.8 &  -00 19            37 &  207 &   23 &   42$\pm$ \phantom{0}4 &   42$\pm$   \phantom{0}4 &    35.5$\pm$ \phantom{0}2.3\\
215.17 & +3.99 &  07 09            42.1 &  -00 09            56 &  189 &   13 &   38$\pm$ \phantom{0}6 &   39$\pm$   \phantom{0}6 &    25.6$\pm$ \phantom{0}5.1\\
215.65 & +3.59 &  07 09            11.3 &  -00 46            29 &  240 &   17 &   32$\pm$ \phantom{0}6 &   32$\pm$   \phantom{0}6 &    31.9$\pm$ \phantom{0}3.7\\
215.73 & -0.70 &  06 54  \phantom{0}2.6 &  -02 48  \phantom{0}3 &  318 &   11 &   62$\pm$           11 &   62$\pm$             12 &    79.6$\pm$ \phantom{0}6.9\\
216.04 & +0.40 &  06 58            32.2 &  -02 34            37 &  268 &    6 &  103$\pm$           16 &  102$\pm$             17 &    70.4$\pm$           10.2\\
216.50 & +1.60 &  07 03            38.5 &  -02 26  \phantom{0}9 &   94 &    9 &    7$\pm$ \phantom{0}5 &    6$\pm$   \phantom{0}5 &     4.8$\pm$ \phantom{0}9.6\\
216.82 & -4.00 &  06 44            13.9 &  -05 16            36 &  210 &   28 &  153$\pm$ \phantom{0}3 &  152$\pm$   \phantom{0}3 &   150.4$\pm$ \phantom{0}3.0\\
216.89 & +0.18 &  06 59            18.3 &  -03 25            51 &  164 &   10 &   88$\pm$ \phantom{0}7 &   87$\pm$   \phantom{0}7 &      --                    \\
217.27 & +4.23 &  07 14            24.2 &  -01 55            29 &  191 &   11 &   64$\pm$ \phantom{0}7 &   65$\pm$   \phantom{0}8 &    59.0$\pm$ \phantom{0}8.0\\
217.45 & +4.78 &  07 16            42.1 &  -01 49            21 &   65 &    8 &  122$\pm$ \phantom{0}5 &  122$\pm$   \phantom{0}5 &   103.0$\pm$           11.5\\
217.48 & -1.37 &  06 54            50.4 &  -04 40  \phantom{0}2 &   68 &    6 &  202$\pm$ \phantom{0}7 &  202$\pm$   \phantom{0}7 &      --                    \\
217.78 & -2.95 &  06 49            45.6 &  -05 39            22 &  465 &   20 &   88$\pm$ \phantom{0}9 &   88$\pm$   \phantom{0}9 &    74.0$\pm$ \phantom{0}5.3\\
217.96 & +4.26 &  07 15            47.0 &  -02 31  \phantom{0}1 & 1410 &   47 &  110$\pm$           11 &  110$\pm$             12 &   124.9$\pm$ \phantom{0}2.3\\
218.01 & -3.95 &  06 46            34.7 &  -06 18            34 &  196 &   12 &   78$\pm$ \phantom{0}7 &   78$\pm$   \phantom{0}7 &    69.5$\pm$ \phantom{0}5.6\\
218.14 & +3.78 &  07 14            23.8 &  -02 54  \phantom{0}9 &  172 &    9 &   88$\pm$ \phantom{0}8 &   89$\pm$   \phantom{0}8 &    98.7$\pm$           10.9\\
218.41 & +2.50 &  07 10            20.1 &  -03 43            26 & 1158 &   27 &  116$\pm$           16 &  115$\pm$             17 &   135.7$\pm$ \phantom{0}1.6\\
218.81 & +0.52 &  07 04  \phantom{0}2.7 &  -04 59            22 &  147 &    6 &   88$\pm$           10 &   88$\pm$             10 &    90.8$\pm$ \phantom{0}9.8\\
219.19 & -4.89 &  06 45            20.1 &  -07 47  \phantom{0}5 &  107 &   10 &   86$\pm$ \phantom{0}5 &   87$\pm$   \phantom{0}5 &    70.6$\pm$ \phantom{0}8.2\\
219.27 & +0.86 &  07 06  \phantom{0}6.4 &  -05 14            25 &  262 &   16 &  103$\pm$ \phantom{0}7 &  103$\pm$   \phantom{0}7 &    95.2$\pm$ \phantom{0}5.3\\
219.48 & +2.79 &  07 13            20.3 &  -04 32            27 &  304 &   15 &   74$\pm$ \phantom{0}8 &   74$\pm$   \phantom{0}8 &    67.8$\pm$ \phantom{0}5.0\\
219.60 & +3.04 &  07 14            28.5 &  -04 31            30 &  235 &    9 &   87$\pm$           10 &   87$\pm$             10 &    95.6$\pm$ \phantom{0}7.8\\
219.75 & -3.63 &  06 50            53.0 &  -07 42            56 &  396 &    4 &   56$\pm$           33 &   56$\pm$             36 &    29.4$\pm$ \phantom{0}6.7\\
220.02 & +1.34 &  07 09            11.9 &  -05 40            59 &  165 &   30 &   69$\pm$ \phantom{0}3 &   69$\pm$   \phantom{0}3 &    56.3$\pm$ \phantom{0}3.6\\
220.07 & -4.70 &  06 47            38.0 &  -08 29  \phantom{0}1 &  187 &    9 &   69$\pm$ \phantom{0}8 &   69$\pm$   \phantom{0}9 &    44.2$\pm$ \phantom{0}7.7\\
220.11 & -4.34 &  06 48            59.5 &  -08 21            33 &  272 &   13 &   78$\pm$ \phantom{0}8 &   78$\pm$   \phantom{0}9 &    78.7$\pm$ \phantom{0}4.9\\
220.35 & -3.93 &  06 50            55.0 &  -08 22            55 &  137 &    6 &   55$\pm$           10 &   56$\pm$             10 &      --                    \\
220.35 & -3.93 &  06 50            55.0 &  -08 22            55 &  132 &    5 &   68$\pm$           10 &   69$\pm$             11 &      --                    \\
220.42 & -4.95 &  06 47            20.0 &  -08 54            34 &  140 &    8 &   90$\pm$ \phantom{0}7 &   92$\pm$   \phantom{0}8 &    91.4$\pm$           10.8\\
220.44 & -0.31 &  07 04  \phantom{0}5.3 &  -06 49  \phantom{0}2 &  158 &    9 &  118$\pm$ \phantom{0}8 &  119$\pm$   \phantom{0}8 &   116.7$\pm$           10.1\\
220.65 & +1.53 &  07 11  \phantom{0}2.2 &  -06 09            22 &  333 &   16 &   94$\pm$ \phantom{0}8 &   94$\pm$   \phantom{0}8 &    99.8$\pm$           10.0\\
220.79 & +0.93 &  07 09            10.2 &  -06 33            30 &  175 &    6 &   86$\pm$           11 &   86$\pm$             12 &      --                    \\
\enddata
\end{deluxetable}

\addtocounter{table}{-1}
\begin{deluxetable}{rrrrrrrrr}
\tablewidth{0 pt}
\tabletypesize{\small}
\tablecolumns{9}
\tablecaption{-- {\sl{Continued}}}
\tablehead{\colhead{$l$} &
\colhead{$b$} & \colhead{$\alpha$} & \colhead{$\delta$} & \colhead{Stokes I$^{\dagger}$} & \colhead{PI$^{\dagger}$} &
\colhead{RM$^{\ddagger}$} &\colhead{RM synthesis} & \colhead{Taylor
RM$^{\ast}$}  \\
\colhead{[\degr]} & \colhead{[\degr]} & \colhead{[h m s]}  & \colhead{[\degr ~  \arcmin ~ \arcsec]} & \colhead{ [mJy]} &
\colhead{[mJy]} & \colhead{[rad m$^{-2}$]} & \colhead{[rad
m$^{-2}$]} & \colhead{[rad m$^{-2}$]}}
\startdata
221.00 & -4.66 &  06 49            27.8 &  -09 17            54 &  200 &   20 &   68$\pm$ \phantom{0}4 &   68$\pm$   \phantom{0}5 &    62.9$\pm$ \phantom{0}7.2\\
221.01 & -2.36 &  06 57            46.2 &  -08 15            19 &  367 &   28 &   78$\pm$ \phantom{0}5 &   78$\pm$   \phantom{0}5 &    75.7$\pm$ \phantom{0}2.8\\
221.06 & -2.95 &  06 55            43.2 &  -08 34            44 &  108 &    6 &   99$\pm$ \phantom{0}8 &   98$\pm$   \phantom{0}8 &      --                    \\
221.32 & +0.84 &  07 09            50.1 &  -07 04            21 &  204 &    7 &  145$\pm$           12 &  145$\pm$             12 &   115.2$\pm$           11.4\\
221.54 & -2.04 &  06 59            53.0 &  -08 35            21 &  142 &    5 &   75$\pm$           11 &   72$\pm$             11 &    81.9$\pm$           15.1\\
222.39 & +4.76 &  07 25            47.6 &  -06 11            46 &  220 &    8 &   22$\pm$           12 &   23$\pm$             12 &    24.1$\pm$ \phantom{0}9.8\\
222.72 & -1.74 &  07 03  \phantom{0}8.6 &  -09 29            56 &  162 &    7 &  164$\pm$           10 &  162$\pm$             10 &   143.6$\pm$           10.5\\
223.07 & -4.99 &  06 52  \phantom{0}1.3 &  -11 17  \phantom{0}6 &  142 &    5 &   73$\pm$           12 &   72$\pm$             12 &    55.9$\pm$           12.2\\
223.25 & +1.35 &  07 15            14.9 &  -08 32            55 &   89 &    7 & -128$\pm$ \phantom{0}6 & -129$\pm$   \phantom{0}7 &      --                    \\
223.27 & +1.10 &  07 14            24.4 &  -08 40            53 &  113 &    7 & -180$\pm$ \phantom{0}8 & -180$\pm$   \phantom{0}8 &  -140.4$\pm$           15.7\\
223.33 & -2.21 &  07 02            35.7 &  -10 15  \phantom{0}5 &  239 &   11 &   67$\pm$ \phantom{0}9 &   67$\pm$   \phantom{0}9 &    54.9$\pm$ \phantom{0}9.2\\
223.54 & -2.82 &  07 00            46.7 &  -10 43            23 &  402 &   22 &   64$\pm$ \phantom{0}7 &   64$\pm$   \phantom{0}7 &    51.0$\pm$ \phantom{0}3.5\\
223.67 & -3.25 &  06 59            28.8 &  -11 01            50 &  110 &    5 &   95$\pm$           10 &   96$\pm$             10 &    48.8$\pm$           17.1\\
224.21 & +2.65 &  07 21            44.1 &  -08 47            13 &  268 &    8 &  -83$\pm$           13 &  -83$\pm$             13 &   -83.8$\pm$ \phantom{0}8.6\\
224.98 & +2.73 &  07 23            28.5 &  -09 25            55 &  317 &   11 & -173$\pm$           11 & -173$\pm$             11 &  -106.9$\pm$ \phantom{0}8.7\\
225.41 & +1.07 &  07 18            21.0 &  -10 35  \phantom{0}1 &  173 &   27 & -547$\pm$ \phantom{0}3 & -276$\pm$   \phantom{0}3 &   127.5$\pm$ \phantom{0}4.6\\
225.75 & +4.94 &  07 32            50.6 &  -09 03            18 &   82 &   10 &  -31$\pm$ \phantom{0}4 &  -31$\pm$   \phantom{0}4 &   -37.5$\pm$ \phantom{0}7.1\\
226.15 & +4.69 &  07 32            42.4 &  -09 31            51 &  215 &   10 &  -31$\pm$ \phantom{0}8 &  -32$\pm$   \phantom{0}9 &   -26.8$\pm$ \phantom{0}7.0\\
228.38 & +1.26 &  07 24            44.0 &  -13 06            46 &  204 &   12 &  -90$\pm$ \phantom{0}7 &  -91$\pm$   \phantom{0}7 &   -83.4$\pm$ \phantom{0}9.0\\
229.12 & +1.18 &  07 25            52.7 &  -13 48            33 &  189 &    7 &  -88$\pm$           11 &  -87$\pm$             12 &   -90.3$\pm$           14.2\\
229.87 & -0.64 &  07 20            43.6 &  -15 19            40 &  684 &   43 &  217$\pm$ \phantom{0}6 &  217$\pm$   \phantom{0}6 &   190.3$\pm$ \phantom{0}2.3\\
230.04 & +3.34 &  07 35            29.9 &  -13 34            50 &  145 &    7 &   81$\pm$ \phantom{0}9 &   80$\pm$   \phantom{0}9 &    50.2$\pm$ \phantom{0}7.8\\
230.09 & -0.62 &  07 21            13.5 &  -15 30            39 &  548 &   30 &  113$\pm$ \phantom{0}7 &  113$\pm$   \phantom{0}7 &    66.9$\pm$ \phantom{0}2.4\\
230.17 & +1.07 &  07 27            32.5 &  -14 46            53 &  281 &   12 &  105$\pm$ \phantom{0}9 &  104$\pm$   \phantom{0}9 &    78.2$\pm$ \phantom{0}6.7\\
231.08 & +4.21 &  07 40            42.2 &  -14 03            46 &  355 &   10 &   50$\pm$           13 &   51$\pm$             14 &    26.1$\pm$ \phantom{0}5.1\\
231.20 & +1.16 &  07 29            55.5 &  -15 38            27 &  214 &   14 &   81$\pm$ \phantom{0}7 &   81$\pm$   \phantom{0}7 &    80.6$\pm$ \phantom{0}6.9\\
231.66 & -0.01 &  07 26            33.3 &  -16 36            26 &  121 &    8 &  188$\pm$ \phantom{0}7 &  188$\pm$   \phantom{0}7 &   169.4$\pm$           10.8\\
231.99 & +2.75 &  07 37            16.2 &  -15 34  \phantom{0}6 &   90 &    9 &   87$\pm$ \phantom{0}5 &   86$\pm$   \phantom{0}5 &    45.9$\pm$ \phantom{0}5.4\\
232.66 & +1.87 &  07 35            25.9 &  -16 34            48 &  253 &   12 &   77$\pm$ \phantom{0}8 &   77$\pm$   \phantom{0}8 &    42.3$\pm$ \phantom{0}6.1\\
232.83 & +3.13 &  07 40            21.2 &  -16 06            32 &  266 &   16 &   58$\pm$ \phantom{0}7 &   58$\pm$   \phantom{0}7 &    58.5$\pm$ \phantom{0}5.0\\
233.39 & -2.60 &  07 20            22.1 &  -19 21            12 &  175 &    7 &   98$\pm$           10 &   98$\pm$             10 &    88.0$\pm$           11.0\\
233.53 & -1.31 &  07 25            28.9 &  -18 52  \phantom{0}6 &  264 &   17 &  148$\pm$ \phantom{0}6 &  148$\pm$   \phantom{0}6 &   136.7$\pm$ \phantom{0}3.9\\
234.75 & +0.87 &  07 36  \phantom{0}2.4 &  -18 53  \phantom{0}9 &  319 &   18 &  174$\pm$ \phantom{0}7 &  174$\pm$   \phantom{0}7 &   156.6$\pm$ \phantom{0}3.3\\
235.28 & -0.43 &  07 32            16.4 &  -19 58            58 &  144 &   16 &  142$\pm$ \phantom{0}4 &  142$\pm$   \phantom{0}4 &   112.1$\pm$ \phantom{0}4.9\\
237.10 & +1.30 &  07 42            32.1 &  -20 43            41 &  153 &   10 &   97$\pm$ \phantom{0}7 &   98$\pm$   \phantom{0}7 &    90.0$\pm$ \phantom{0}8.3\\
237.51 & +0.17 &  07 39            12.5 &  -21 38            44 &  147 &    9 &   69$\pm$ \phantom{0}7 &   69$\pm$   \phantom{0}7 &    64.7$\pm$ \phantom{0}8.8\\
237.72 & +4.34 &  07 55  \phantom{0}2.3 &  -19 43            27 &  258 &   29 &  264$\pm$ \phantom{0}4 &  264$\pm$   \phantom{0}4 &   233.6$\pm$ \phantom{0}2.7\\
240.53 & -3.57 &  07 31            16.9 &  -26 06  \phantom{0}0 &  139 &    8 &  258$\pm$ \phantom{0}8 &  259$\pm$   \phantom{0}8 &   235.3$\pm$           12.8\\
241.70 & -1.87 &  07 40            25.7 &  -26 17            15 &  257 &   12 &  203$\pm$ \phantom{0}9 &  205$\pm$   \phantom{0}9 &   171.8$\pm$ \phantom{0}6.0\\
242.17 & +4.39 &  08 05            12.4 &  -23 28            55 &   99 &    6 &   98$\pm$ \phantom{0}8 &   99$\pm$   \phantom{0}9 &    85.4$\pm$           12.4\\
243.61 & +3.38 &  08 04            49.2 &  -25 13            55 &  153 &   11 &   96$\pm$ \phantom{0}6 &   96$\pm$   \phantom{0}6 &    58.1$\pm$           10.2\\
245.43 & +1.83 &  08 03            16.0 &  -27 36            18 &  138 &   12 &   99$\pm$ \phantom{0}5 &   98$\pm$   \phantom{0}5 &    77.6$\pm$ \phantom{0}5.2\\
245.80 & +2.01 &  08 04            51.5 &  -27 49            11 &  720 &   26 &   81$\pm$           10 &   81$\pm$             11 &    79.0$\pm$ \phantom{0}2.5\\
246.98 & +2.78 &  08 10            41.0 &  -28 23            40 &  155 &    6 &  133$\pm$           11 &  133$\pm$             12 &    64.1$\pm$           14.0\\
247.06 & -4.73 &  07 41  \phantom{0}4.7 &  -32 21            45 &  161 &   10 &  189$\pm$ \phantom{0}7 &  188$\pm$   \phantom{0}7 &   168.9$\pm$           13.6\\
247.12 & +1.27 &  08 05            15.8 &  -29 19            50 &  117 &    7 &   93$\pm$ \phantom{0}7 &   92$\pm$   \phantom{0}8 &    62.9$\pm$ \phantom{0}9.7\\
247.13 & +3.14 &  08 12            25.9 &  -28 19            40 &  326 &   18 &  102$\pm$ \phantom{0}7 &  101$\pm$   \phantom{0}7 &    99.2$\pm$ \phantom{0}4.3\\
247.25 & -1.94 &  07 52            52.3 &  -31 07            13 &  421 &   16 &  134$\pm$           10 &  135$\pm$             11 &    94.8$\pm$ \phantom{0}4.9\\
249.81 & -1.92 &  07 59            14.3 &  -33 18  \phantom{0}0 &  233 &   12 &  163$\pm$ \phantom{0}8 &  163$\pm$   \phantom{0}8 &   156.3$\pm$ \phantom{0}8.4\\
251.59 & -2.17 &  08 02            44.0 &  -34 56            20 &  118 &    6 &  157$\pm$ \phantom{0}8 &  157$\pm$   \phantom{0}9 &   136.8$\pm$           12.4\\
\enddata
\label{table:1}
\tablecomments{$\dagger$ Both Stokes I and PI (Linear polarized intensity) are determined as the average over all 14 channels for the same pixel, being that with
the highest PI.  \\
\hspace{15mm} {$\ddagger$} RM calculations are done using the angle fitting algorithm described in the text. \\
\hspace{15mm} {$\ast$} Missing values do not have a corresponding
source within 1\arcmin.\\
 }
\end{deluxetable}

\singlespace


%
%

\begin{center}

\begin{table}
\begin{center}
\caption{Best fit  B-field values for our model sectors shown in Figure \ref{fig:Model_diagrams}.\label{table:2}}
\end{center}
\begin{tabular}{| r | r | r | r |}
\hline
\hline
{\bf{Sector and Region}} & {\bf{Magnetic pitch angle}}  & {\bf{Radial dependence}} & $\mathbf{B}^{ \dagger}$  {\bf{(in $\mu$G)}} \\
\hline
\hline
\textbf{Sector A}	&	& 	&	\\
1       &       0$^\circ$ & $R^{-1}$ & -1.20$\pm$0.48\\
\hline
\textbf{Sector B}	&	& 	&	\\
1	&	0$^\circ$ & constant	& 0.85$\pm0.06$\\
2	&	11.5$^\circ$ & constant	& 1.0$\pm0.4\phantom{0}$\\
3       &       11.5$^\circ$ &  constant        & $-0.54\pm0.07$\\
4       &       11.5$^\circ$ &  constant        & $-0.92\pm0.06$\\
5       &       11.5$^\circ$ & constant         & $1.71\pm0.06$\\
6       &       11.5$^\circ$ &  constant        & $-0.90\pm0.08$\\
7       &       11.5$^\circ$ &  constant        & $-0.34\pm0.06$\\
8       &       0$^\circ$ & $R^{-1}$ & $-0.78\pm0.12$\\
\hline
\textbf{Sector C}	&	& 	&	\\
1       &       0$^\circ$    & constant         & $-0.15\pm0.04$\\
2       &       11.5$^\circ$ &  constant        & $-0.40\pm0.01$\\
3       &       11.5$^\circ$ &  constant        & $2.23\pm0.13$\\
4       &       11.5$^\circ$ &  constant        & $0.09\pm0.05$\\
5       &       0$^\circ$    & $R^{-1}$            & $-0.86\pm0.09$\\
\hline
\end{tabular}
\vspace{4mm}
\tablecomments{{$\dagger$} Indicated field strengths for $R^{-1}$ regions are values at a Galactocentric radius \\
 of 8.5 kpc (ie.\ at the Sun.)  Positive values correspond to a counter-clockwise field as viewed \\ 
 from  the north Galactic pole, while negative values correspond to a clockwise field.}

\end{table}

\end{center}

\end{document}